\documentclass[a4paper, 11pt]{article}

\usepackage{jheppub}

\usepackage{graphicx,wrapfig,float,slashed,subcaption,bm}
\usepackage{amsmath,amssymb,epsfig,graphicx,xcolor}
\usepackage{epstopdf}
\usepackage{booktabs}
\usepackage{ragged2e}
\usepackage{verbatim}
\usepackage{makecell}
\usepackage{multirow}
\usepackage{slashed}
\usepackage{nameref}

\usepackage{arydshln}
\usepackage{tabu}
\usepackage{diagbox}
\newlength\replength
\newcommand\repfrac{.33}

\newcommand\rulewidth{.6pt}
\newcommand\tdashfill[1][\repfrac]{\cleaders\hbox to \replength{%
  \smash{\rule[\arraystretch\ht\strutbox]{\repfrac\replength}{\rulewidth}}}\hfill}
\newcommand\tabdashline{%
  \makebox[0pt][r]{\makebox[\tabcolsep]{\tdashfill\hfil}}\tdashfill\hfil%
  \makebox[0pt][l]{\makebox[\tabcolsep]{\tdashfill\hfil}}%
  \\[-\arraystretch\dimexpr\ht\strutbox+\dp\strutbox\relax]%
}
\newcommand\tdotfill[1][\repfrac]{\cleaders\hbox to \replength{%
  \smash{\raisebox{\arraystretch\dimexpr\ht\strutbox-.1ex\relax}{.}}}\hfill}

\newcommand{\nc}{\newcommand}
\nc{\non}{\nonumber}
\nc{\hc}{\hbox {H.c.}}
\nc{\noi}{\noindent}
\nc{\barx}{\bar{x}}
\nc{\pbarn}{\;\hbox {pb}}
\nc{\fbarn}{\;\hbox {fb}}

\nc{\hsp}{\hspace{0.5cm}}
\nc{\lsp}{\hspace{1cm}}
\nc{\Lsp}{\hspace{2cm}}
\nc{\LLsp}{\lsp\lsp}
\nc{\lra}{\longrightarrow}
\nc{\p}{\prime}
\nc{\sgn}{\text{sgn}}
\nc{\ph}{\varphi}
\nc{\op}{{\cal O}}

\nc{\beq}{\begin{equation}}  \nc{\eeq}{\end{equation}}
\nc{\bea}{\begin{eqnarray}}  \nc{\eea}{\end{eqnarray}}
\nc{\baa}{\begin{array}}     \nc{\eaa}{\end{array}}
\nc{\bit}{\begin{itemize}}   \nc{\eit}{\end{itemize}}
\nc{\ben}{\begin{enumerate}} \nc{\een}{\end{enumerate}}
\nc{\bce}{\begin{center}}    \nc{\ece}{\end{center}}
\nc{\bpm}{\begin{pmatrix}}   \nc{\epm}{\end{pmatrix}}
\nc{\bvt}{\begin{verbatim}}  \nc{\evt}{\end{verbatim}}

\def\lsim{\mathrel{\raise.3ex\hbox{$<$\kern-.75em\lower1ex\hbox{$\sim$}}}}
\def\gsim{\mathrel{\raise.3ex\hbox{$>$\kern-.75em\lower1ex\hbox{$\sim$}}}}

\def\udots{\mathinner{\mkern1mu\raise1pt\vbox{\kern7pt\hbox{.}}\mkern2mu\raise4pt\hbox{.}\mkern2mu\raise7pt\hbox{.}\mkern1mu}}

\def\mev{\;\hbox{MeV}}
\def\gev{\;\hbox{GeV}}
\def\tev{\;\hbox{TeV}}
\def\ns{\;\hbox{ns}}

\newcommand{\met}{E^\mathrm{miss}_T}

\allowdisplaybreaks

\newcommand\fverb{\setbox\fverbbox=\hbox\bgroup\verb}
\newcommand\fverbdo{\egroup\medskip\noindent%
			\fbox{\unhbox\fverbbox}\ }
\newcommand\fverbit{\egroup\item[\fbox{\unhbox\fverbbox}]}
\newbox\fverbbox

\preprint{\begin{flushright}
    UTWI-14-2022\\
    RBI-ThPhys-2022-41
\end{flushright}}

\title{Optimizing pixel tracklet searches for shorter lifetimes}

\author[a]{Matthew Gignac}
\author[b]{Can Kilic}
\author[c]{Rakhi Mahbubani}
\author[b]{Taewook Youn}
\affiliation[a]{University of California Santa Cruz, Santa Cruz CA, USA}
\affiliation[b]{Weinberg Institute, Department of Physics, University of Texas at Austin,
  Austin, TX 78712, USA}
\affiliation[c]{Rudjer Boskovic Institute, Division of Theoretical Physics, Bijenička 54, HR-10000 Zagreb, Croatia}
\emailAdd{matthew.gignac@cern.ch}
\emailAdd{kilic@physics.utexas.edu}
\emailAdd{rakhi@cern.ch}
\emailAdd{taewook.youn@utexas.edu}

\begin{document}

\abstract{Pixel tracklets, disappearing tracks reconstructed with only pixel hits, have proven to be a promising technique in LHC analyses to search for dark matter candidates at the LHC that belong to a nearly-degenerate electroweak multiplet. However, a Pseudo-Dirac electroweak doublet fermion, arguably the most interesting such possibility, has a shorter lifetime and therefore existing tracklet searches are less sensitive in this case. We assess the performance of a tracklet search optimized for shorter lifetimes by requiring only three pixel hits for the tracklet reconstruction, and by demanding an accompanying soft track for suppressing backgrounds. We estimate how far the sensitivity of existing searches can be extended into the region of parameter space with this optimized search.
}


\maketitle
\flushbottom
\section{Introduction}

While the Standard Model (SM) of particle physics has proven to be extremely successful in describing nature at short distances, there are a number of fundamental open questions that it cannot address, such as the existence of dark matter (DM), and the naturalness of the electroweak symmetry breaking scale. These two important puzzles may in fact be connected: Firstly, the introduction of partner particles to address the naturalness puzzle often requires the introduction of a $Z_2$ symmetry which renders the lightest partner particle stable and therefore a good DM candidate. Secondly, a thermal relic with a TeV scale mass and an electroweak interaction cross section would obtain a relic abundance consistent with astrophysical observations, often referred to as the WIMP miracle.

Unfortunately, the most motivated models where the DM particle belongs to a nontrivial multiplet of the weak $SU(2)$ group, a weakly-interacting massive particle (WIMP) in the original usage of the word, are severely constrained by indirect detection searches~\cite{Cohen:2013ama,Fan:2013faa}. There is however one notable scenario that is still viable: a Pseudo-Dirac electroweak-doublet fermion with hypercharge $1/2$, which in a supersymmetric setting would correspond to a higgsino. While a pure higgsino state accounting for all of the measured relic abundance is ruled out by direct detection, a small amount of mixing with an electroweak singlet, such as a bino can induce an $\mathcal{O}(\textrm{keV})$ splitting between the two neutral Majorana states.  This is sufficient to evade direct detection bounds~\cite{Tucker-Smith:2001myb} without significantly altering the cosmology or collider phenomenology of the higgsino, and is arguably the best-motivated case for a viable minimal WIMP model~\cite{Krall:2017xij}.

It is well known that after electroweak symmetry breaking, the components of an electroweak multiplet acquire mass differences of order $\alpha \mkern 1mu m_Z$, with the charged states becoming heavier than the neutral one~\cite{Mizuta:1992ja,Thomas:1998wy}. Since the involved mass differences are quite small, the charged states are long-lived for collider purposes. When they are produced in collisions, they travel a microscopic distance before they decay to the neutral component, which is a DM candidate and thus unobservable in detectors, and additional particles - in most scenarios a single pion or a lepton-neutrino pair~\cite{Chen:1996ap,Thomas:1998wy,Chen:1999yf}.

Recently, experimental searches relying on `disappearing' charged tracks have proven to be promising in searching for minimal WIMP candidates at the Large Hadron Collider (LHC)~\cite{ATLAS:2022rme,CMS:2020atg}. In these searches, one looks for the charged component decaying within or just outside the inner pixel detector, but not traversing the entire tracker. Such a particle fails the standard track reconstruction algorithms, but may be identified using a more relaxed definition of a tracklet, reconstructed exclusively from hits in the innermost tracker layers, with a veto on hits in outer layers.  In the case of an electroweak triplet with zero hypercharge, corresponding in the supersymmetric case to a pure wino, the mass splitting is only $\sim$160~MeV, resulting in decay lengths that are of order the inner tracker radii, and thus very suitable for tracklet searches. On the other hand, for an electroweak doublet with hypercharge $1/2$, the mass splitting is larger, around 350~MeV, resulting in significantly shorter lifetimes. Such particles can only pass the selection criteria of a tracklet search if they are highly boosted, which means that the sensitivity of the tracklet search is significantly reduced.  Current searches at ATLAS and CMS, which require at least 4 inner tracker hits, exclude pure doublet WIMPs of mass 210 GeV at 95\% C.L. (ATLAS) and $~$160 GeV (CMS), leaving the bulk of the doublet parameter space still unconstrained by LHC searches.

Since the limiting factor is the shorter lifetime in the case of the doublet, we aim in this paper to optimize a shorter tracklet search along the lines proposed in ref.~\cite{Mahbubani:2017gjh, Fukuda:2017jmk}, but with careful attention to background modeling, in order to explore a wider region within the parameter space.  We base our analysis on the Run 2 36.1 fb$^{-1}$ ATLAS search~\cite{Aaboud:2017mpt}, using tracklets that leave three or more hits in the pixel detector, which we refer to as a `3-layer tracklet' analysis.\footnote{This result has since been superseded by ref.~\cite{ATLAS:2022rme} with an integrated luminosity of 136 fb$^{-1}$ and modified kinematic selection.  Although we continue to use the 2017 selection cuts in this analysis, we display the limits of the updated analysis where relevant for the purposes of comparison.}   Unsurprisingly, relaxing the tracklet definition also results in a significant increase in the number of background events. In order to suppress the background, we take advantage of the charged pion produced by the decay of the charged state, and require in addition a soft track originating close to the endpoint of the tracklet.  The signal selection efficiencies for this option was studied in ATLAS Note ref.~\cite{ATLAS:2019pjd}.

In the next section, we begin by describing the details of the $36.1$ fb$^{-1}$ ATLAS search, ref.~\cite{Aaboud:2017mpt}, the features of the signal and the dominant backgrounds, and our modeling of both, demonstrating that we can accurately reproduce the ATLAS results. Having thus validated our methods, in section~\ref{sec:newsearch} we then describe the modified tracklet definition that we use to optimize the reach for the electroweak doublet, and we extrapolate the signal and backgrounds, mapping out the region of parameter space that can be probed beyond the existing limits. We conclude in section~\ref{sec:conclusions} and outline promising directions for further study.

\section{Analysis and Validation}

\subsection{Theoretical framework}

We carry out our analysis within a simplified framework where the SM is supplemented with a new weak-multiplet, color-singlet fermion with hypercharge $q_Y$ and mass $m$. The mass term can be Dirac (for $q_Y\ne 0$) as in the case of the doublet, or Majorana (for $q_Y=0$) as in the case of the triplet. The Lagrangian for this simplified model is given by  
\beq
{\mathcal L} = \xi \left(\overline{\chi}i \slashed{D}\chi - m \overline{\chi}\chi\right),
\label{eq:simplifiedmodel}
\eeq
with $\xi=1/2$ for the Majorana case and $\xi=1$ for the Dirac case, and
\beq
D_{\mu}=\partial_{\mu}-i g W_{\mu}^{a} \tau^{a}-i g' q_{Y} B_{\mu}.
\eeq
We use the normalization ${\rm Tr}[\tau^a \tau^b]=\frac12 \delta^{ab}$ for the $SU(2)$ generators.  We assume the new fermion is charged under a $Z_2$ parity, under which the SM is neutral.  This precludes any Yukawa-type mixing between the new fermion and SM leptons. For convenience we will refer to the neutral and charged components as the `neutralino' and `chargino' respectively although our analysis does not rely on any assumptions about the UV origins of this framework, other than assuming a large mass gap between the states of interest and any other BSM states, such that the effects of the latter can be neglected.

All components of the multiplet $\chi$ are mass-degenerate at tree level, but electroweak loops raise the mass of the chargino with respect to the neutralino, giving rise to a splitting $m_{\chi^\pm}-m_{\chi^0} \sim {\mathcal O} \left( \alpha\, m_Z \right)$.  The precise value of this splitting is specific to the quantum numbers of the multiplet in question: for the doublet this ranges from
$\sim 250-350\mev$  at one loop, depending on the doublet mass~\cite{Thomas:1998wy}, which corresponds to a chargino proper lifetime, $\tau_{\chi^\pm}$, of between 20 and 60 ps. The chargino decays predominantly to a neutralino and charged pion~\cite{Chen:1996ap,Chen:1999yf}.  The $Z_2$ parity ensures the stability of the neutralino, which escapes undetected; leaving a rather soft charged pion as the only visible trace of chargino decay at hadron colliders.  Existing analyses thus rely on the presence of a hard jet recoiling against the chargino-neutralino or chargino-chargino system, giving rise to significant missing energy, which can be used to trigger on these events.

\begin{figure}
	\centering
	\includegraphics[width=0.55\textwidth]{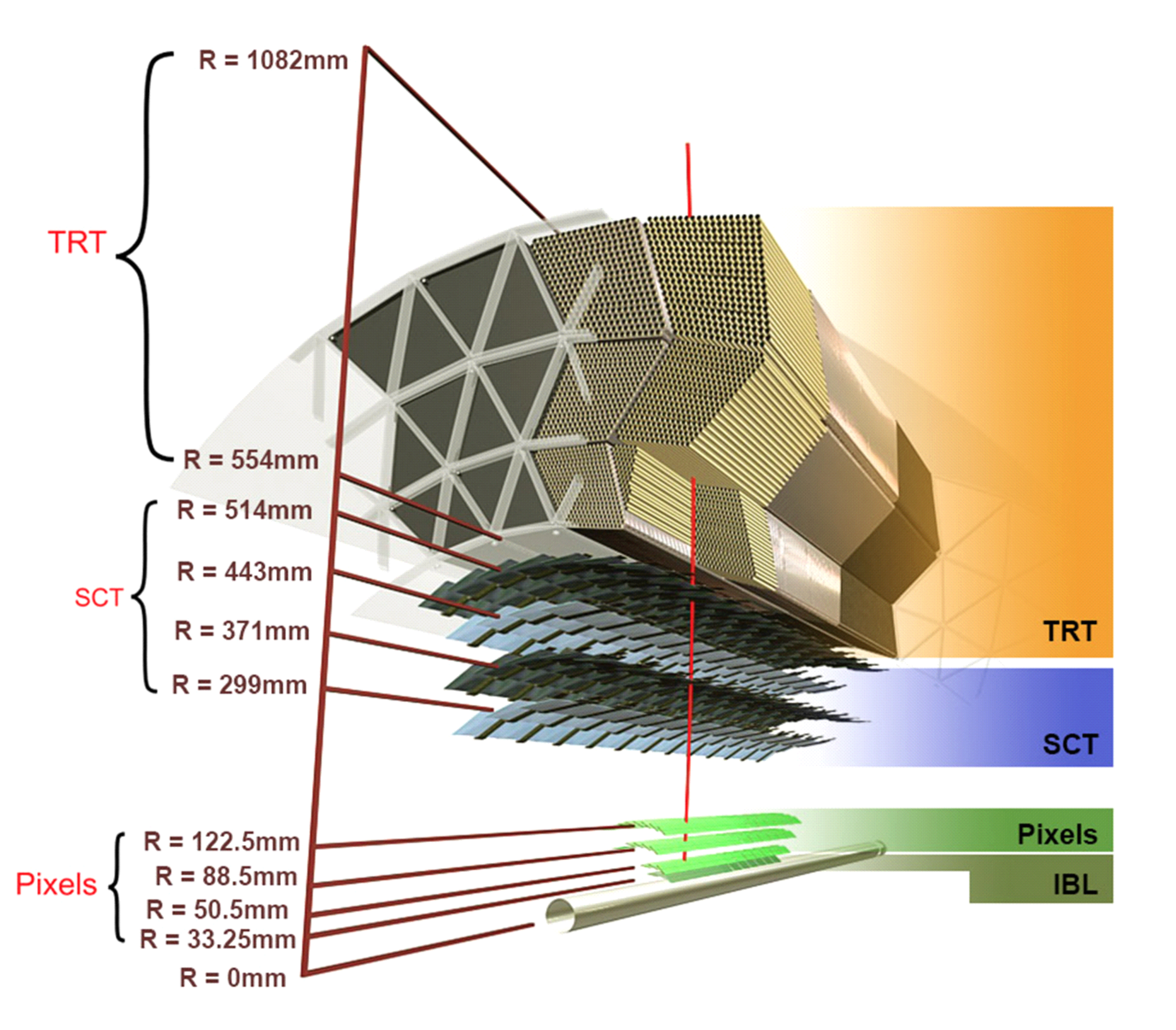}~
        \hspace{0.5in}
	\includegraphics[width=0.25\textwidth]{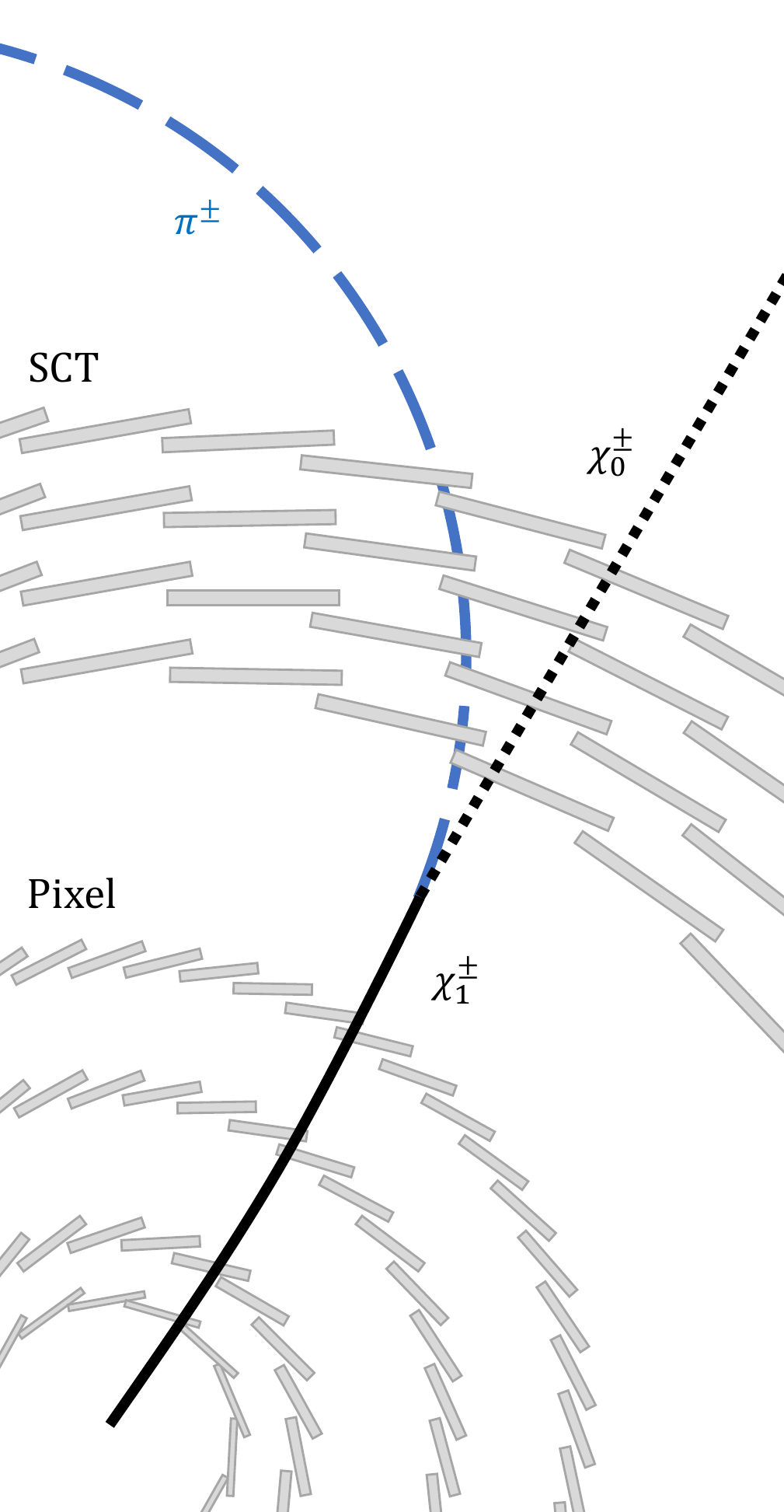}
        \caption{{\it Left:} ATLAS inner tracking detector.  The relevant layers in this analysis are the Insertable B-Layer (IBL) and pixel tracking layers, within which the tracklets are contained.  Tracklets consisting of any hits in the semiconductor tracker (SCT) layers are vetoed.  Figure taken from ref.~\cite{ATLAS:2020ixw}.  {\it Right:} Illustration of a chargino decay giving rise to a 4-hit tracklet.}
        \label{fig:ATLASTracker}
\end{figure}

With their $\mathcal{O}(\mathrm{mm})$ decay length, most charginos from an EW doublet have a high probability of decaying before they traverse even the inner pixel layers of an LHC detector.  However some rare, highly-boosted charginos leave a trace across multiple pixel layers before decaying, resulting in a `disappearing' charged track signature.  These short tracks, or `tracklets', cannot be reconstructed by standard charged-track reconstruction algorithms.  Instead a secondary algorithm needs to be run on the remaining pixel hits after conventional tracks, which extend beyond the pixel detector, have been reconstructed.  
See figure~\ref{fig:ATLASTracker} for the geometry and layout of the ATLAS inner tracking detector, and an illustration of a chargino decay giving rise to a 4-hit tracklet.

SM backgrounds to this process consist of two main components.  The first is due to long-lived charged particles (electrons, muons, charged hadrons) that undergo a large deflection before reaching the silicon layers due to interaction with the detector material, or brehmstrahlung for electrons. The second component is combinatorial fakes, hits from unrelated charged tracks, dominantly originating from pile-up vertices, that are accidentally reconstructed as a tracklet.

\subsection{Signal Modeling}
\label{sec:signal}

\begin{figure}
	\centering
	\includegraphics[width=0.45\textwidth]{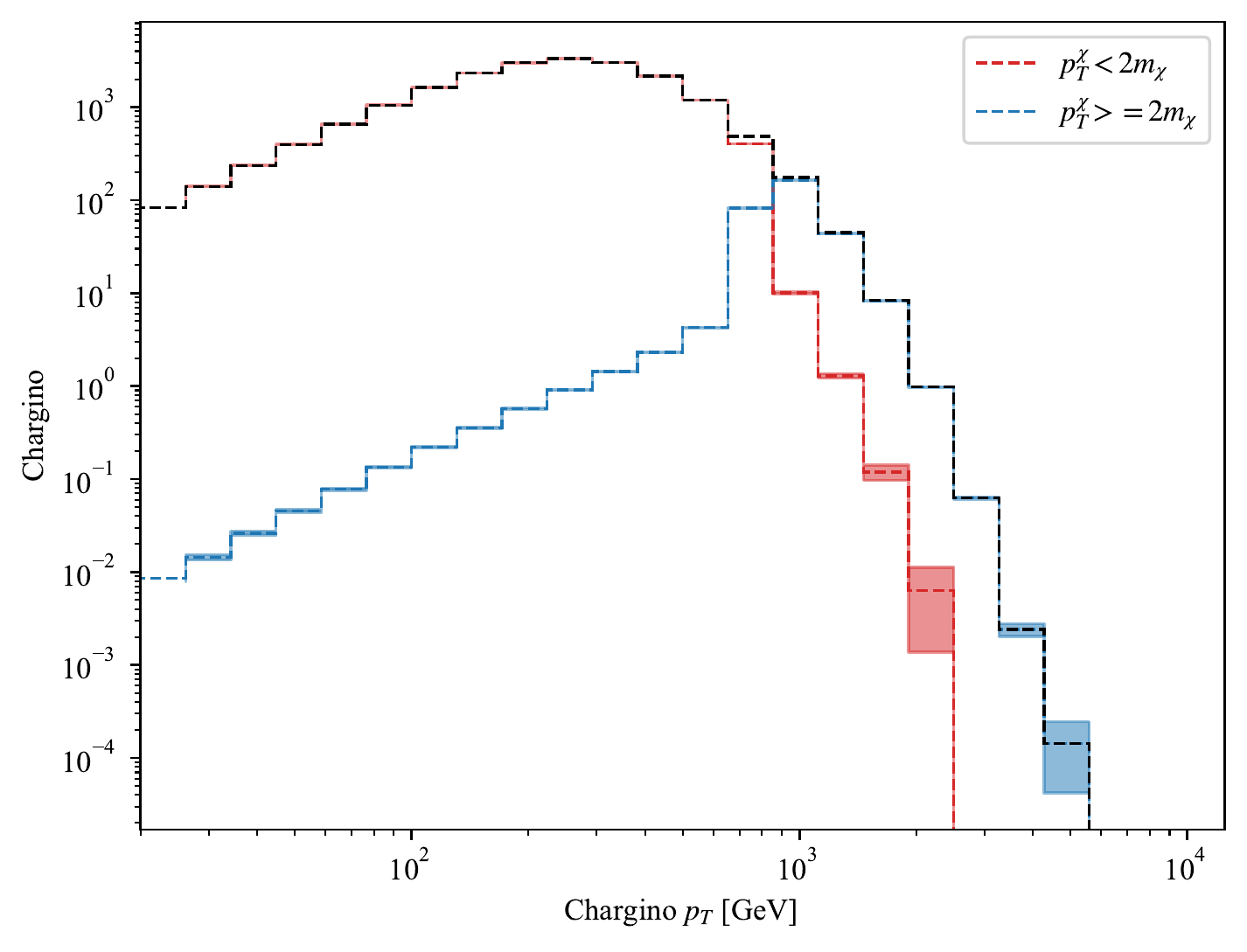}~
        \hspace{0.5in}
	\includegraphics[width=0.45\textwidth]{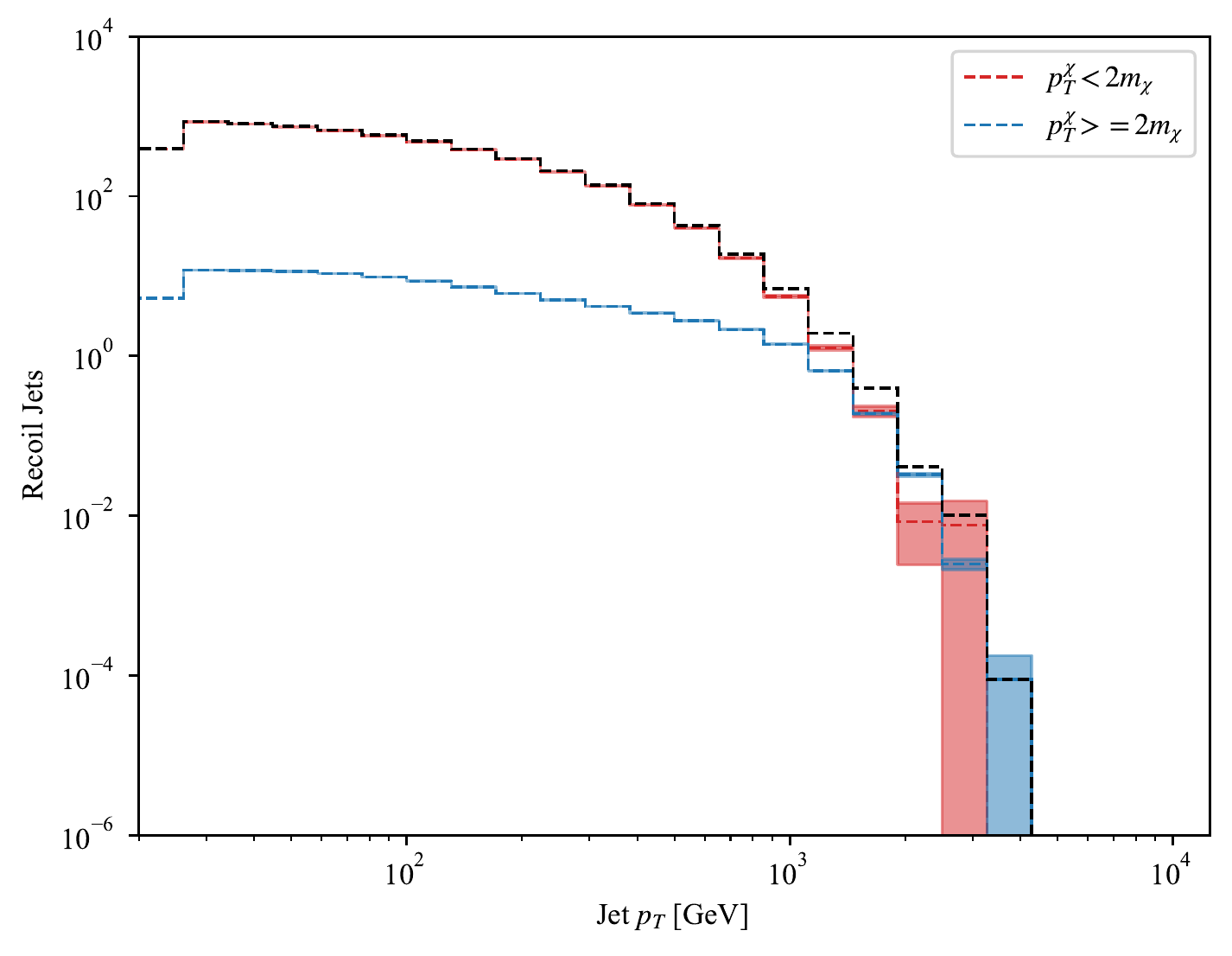}
        \caption{The $p_T$ spectrum of the chargino (left panel, truth level) and hardest jet (right panel, including detector energy resolution), where we show separately the contributions and the uncertainties of the two samples generated with $p_T < 2 m_{\chi^\pm}$ (in red) and  $p_T \ge 2 m_{\chi^\pm}$ (in blue) at the generator level, and combined to get the total spectrum (black dashed). In the left panel, the high $p_T$ sample dominates the chargino $p_T$ spectrum above 1~TeV, whereas the jet $p_T$ spectrum is not well correlated with the chargino $p_T$, and therefore both $p_T$ samples contribute even at the highest bins of the right panel.}
        \label{fig:2binspectra}
\end{figure}

In our Monte Carlo modeling of signal, chargino-neutralino pairs and chargino-chargino pairs are produced at leading order through Drell-Yan processes, merged and matched up to 2 additional partons using MadGraph5\_aMC@NLO 2.7.3~\cite{Alwall:2014hca}, with $p_T > 10$~GeV. Parton-jet matching is done using the CKKW-L merging scheme~\cite{Lonnblad:2011xx} with matching scale set to $m_{\chi^\pm}/4$.  When necessary to obtain a statistically sufficient number of boosted charginos, we generate events in two separate $p_T$ bins: $p_T < 2 m_{\chi^\pm}$ and  $p_T \ge 2 m_{\chi^\pm}$, with merging scales $m_{\chi^\pm_1} / 4$ and $\sqrt 5m_{\chi^\pm_1} / 4$ respectively.  Our results are robust to the choice of scale, and we show in figure~\ref{fig:2binspectra} the resulting $p_{T}$ spectra of the chargino and the hardest jet in the event.  We use the pre-defined MSSM\_SLHA2 UFO model~\cite{Degrande:2011ua, Duhr:2011se} for the purposes of simulation, manually decoupling all supersymmetry (SUSY) particles in the parameter card except for the neutralinos and chargino making up the doublet. We rescale the signal cross section to the NLO+NLL values obtained in ref.~\cite{Fuks:2012qx,Fuks:2013vua}.

While our focus is the reach of tracklet searches for a pure EW doublet, in order to remain general and model-independent, we present our results by scanning over the chargino mass (which fixes the kinematics of the event) and lifetime separately, as the latter is determined by the chargino-neutralino mass splitting, which can depend on the specific UV model. We do however mark the curve in the $(m_{\chi}, c\mkern 1mu\tau)$ plane on which the pure EW doublet lies in our plots, where we set the chargino-neutralino splitting in accordance with the pure EW doublet scenario as computed in ref.~\cite{Thomas:1998wy}.  We compute the total lifetime of the doublet using private code based on ref.~\cite{Chen:1996ap}. In chargino decays, we set the branching fraction to pions to one for simplicity.  Charginos are decayed, and events are showered and hadronized using Pythia 8.243~\cite{Sjostrand:2014zea} with multi-parton interactions (MPI) turned off, and the resulting events are piped through Delphes 3.4.2~\cite{deFavereau:2013fsa} for detector simulation, with the default ATLAS detector card, and anti-$k_T$ jet clustering~\cite{Cacciari:2008gp} with radius parameter 0.4.  The NNPDF23LO PDF set~\cite{NNPDF:2014otw} is used throughout.  Since Delphes does not propagate disappearing tracks we have no means of simulating tracklet smearing or reconstruction.  Instead we will manually smear the $p_T$ of the tracklet using smearing functions published by ATLAS, and simply apply an overall efficiency factor in order to account for inefficiencies in the true reconstruction algorithm due to finite tracker resolution, pile-up and MPI.

The trajectory of the chargino and the decay pion are computed using
private code, modeling the inner detector as a perfect cylinder with a
constant solenoidal magnetic field of 2 Tesla. The magnetic field is relevant only in modeling the pion trajectory, while the charginos, which have a much higher mass and therefore a much larger $p_T$ for the same boost factor, have a trajectory that is almost linear.
Later in this study we will include the soft pion track in the signal selection, along the lines explored in ref.~\cite{ATLAS:2019pjd}, in order to suppress backgrounds.
The $p_T$ of a reconstructed track is determined from its radius of
curvature, and for a particle with unit charge the two quantities
are approximately related as:
\[
\frac{p_T}{\mev}=6 \frac{r}{\textrm{cm}}
\]
Determining the $p_T$ of a short reconstructed tracklet is
challenging in practice, particularly at large $p_T$ (or radius of
curvature), as characteristic of chargino tracks.   We include resolution effects by manually smearing the generator-level chargino tracklets using the following function, which was determined in Ref.~\cite{Aaboud:2017mpt,ATLAS:2022rme} to closely approximate the difference, $\Delta$ between the true chargino $q/p_T$ and the smeared `reconstructed' $q/p_T$:

\begin{eqnarray}
  f(z)&=&\left\{ \begin{array}{l c}
      \exp{\left(\alpha\left(z+\alpha/2\right)\right)}, & z < -\alpha \\
      \exp{\left(-z^2/2\right)}, & -\alpha < z < \alpha \\
      \exp{\left(-\alpha\left(z-\alpha/2\right)\right)}, & z > \alpha 
      \end{array}
      \right.\nonumber\\
      \textrm{for} \,\,z&=& \frac{\Delta\left(q/p_T\right)}{\sigma}\,. \nonumber\\
      \label{equ:ptsmearing}
\end{eqnarray}
We take the parameter values $\alpha=1.64,\;\sigma=14.03$ TeV$^{-1}$, corresponding to the high-$p_T$ limit given in Table 4 of ref.~\cite{ATLAS:2022rme}.  The values in this table were extracted by comparing `tracklets' in $Z\to e^+ e^-$ events where the electron is reconstructed using only pixel hits, to simulations.

\subsection{Reconstruction and event selection}

The reconstruction and selection criteria we use are chosen to mimic those in the ATLAS analysis ref.~\cite{Aaboud:2017mpt}.  Jets are reconstructed in Delphes using FastJet~\cite{Cacciari:2008gp,Cacciari:2011ma}, with $R=0.4$ and $p_T>20$ GeV, and required to have $|\eta| < 2.8$.  Leptons are also reconstructed in Delphes with $p_T>10\gev$ and $\eta < 2.47 \;(2.7)$ for electrons (muons).  The Delphes missing transverse momentum  $E_T^\mathrm{miss}$ is computed by summing energy `deposits' in calorimeter towers.  Event and tracklet selection are summarized in table~\ref{tab:EventSelection}

\begin{table}[h!]
  \renewcommand{\arraystretch}{1.2}
  \begin{tabular}{@{}c@{}}
    \toprule
    Zero leptons \\
    At least one jet with $p_T>140$ GeV\\
    $E_T^\mathrm{miss} > 140\gev$ ($90 \gev < E_T^\mathrm{miss} < 140 \gev$) in the high-$\met$ signal (low-$\met$ control) region\\
    $\Delta\phi( E_T^\mathrm{miss},j) > 1.0$ for the four highest-$p_T$ jets in the event with $p_T>50\gev$.\\
\tabucline[0.4pt off 2pt]{-}
           $p_T > 20\gev,\,0.1 \le|\eta|\le 1.9$\\
          $\Delta R (\textrm{tracklet}, j) > 0.4$ for jets with $p_T > 50\gev$\\
Tracklet has transverse length $r\ge 122.5$ mm (to outermost pixel layer)\\
        Tracklet does not extend into the SCT, $r\le299$ mm (`disappearance condition')\\
             
    \bottomrule
  \end{tabular}
  \caption{Event selection (above the dashed line) and tracklet selection (below the dashed line) criteria used in our validation.  The tracklet selection criteria are applied to generator-level charginos.}\label{tab:EventSelection}
\end{table}
The tracklet selection criteria, shown below the dashed line in table~\ref{tab:EventSelection}, are applied to generator-level charginos.  To maximize statistics, we populate our analysis using all charginos in selected events, weighting each one by the probability that it will decay in the radial interval satisfying the final two tracklet selection criteria.  We assume all charginos satisfy the ATLAS `quality requirement', a set of criteria designed to suppress fake tracklets, with 100\% efficiency.  These criteria include constraints on the minimum number and quality of hits constituting a reconstructed tracklet, its transverse and longitudinal impact parameters and its goodness of fit.  Fake backgrounds are discussed in more detail in Section~\ref{sec:fakebg}.    

\begin{figure}
	\begin{center}
		\includegraphics[width=0.65\textwidth]{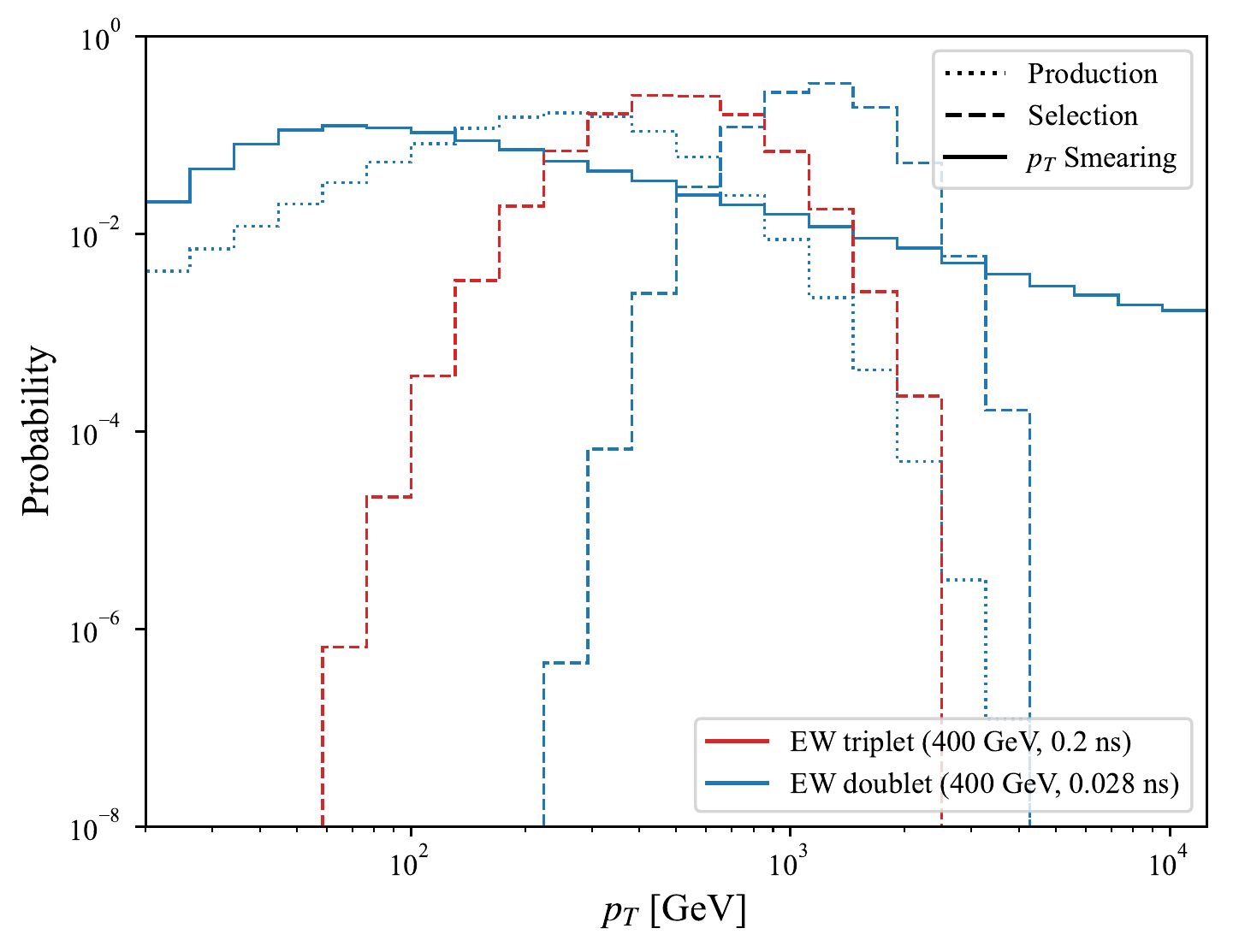}
		\caption{Unit-normalized $p_T$ distributions for a 400-GeV chargino from an EW triplet (red lines) and doublet (blue lines), with proper lifetimes $\tau_{\chi^\pm}=0.2 \mathrm{\ns}$ and $0.028\ns$, corresponding to the pure wino and pure higgsino scenarios, respectively.  The spectra are shown at generator level (dotted lines), after event and tracklet selection (dashed lines) and after smearing (solid lines).  Note the doublet and triplet distributions are indistinguishable at production, confirming that event kinematics are dependent only on the mass of the state, and not its EW quantum numbers. The significant difference in the chargino lifetimes are evident in the distributions after selection, with the shorter-lived doublet requiring a larger transverse boost in order to satisfy the 4-hit requirement on tracklets.  However these differences are obscured once resolution effects due to imperfect tracklet reconstruction are taken into account, and are not fully exploited in the analysis.}
                 \label{fig:pT4}
	\end{center}
\end{figure}

The unit-normalized $p_T$ spectrum for a pure EW triplet with $(m_{\chi^\pm},\tau_{\chi^\pm})=(400\gev, 0.2\ns)$ (red lines) can be compared with that for a pure EW doublet with hypercharge $1/2$ $(m_{\chi^\pm},\tau_{\chi^\pm})=(400\gev,0.028\ns)$ (blue lines) in figure~\ref{fig:pT4}.  The benchmark lifetimes chosen correspond to the pure wino and pure higgsino scenarios respectively.  The triplet parameters were chosen for validation with ATLAS, Ref.~\cite{Aaboud:2017mpt}, with mass spectrum and decay width calculated using ISAJET 7.88~\cite{Paige:2003mg} for minimum anomaly-mediated supersymmetry breaking, with $\tan\beta=5$, $\mu>0$ and the universal scalar mass $m_0= 5\tev$.  The line style in the figure indicates the stage of the analysis at which the spectrum was obtained, with dotted lines corresponding to generator level; dashed lines to spectra after event- and tracklet selection, and solid lines to the `reconstructed' (smeared) $p_T$ distributions.  Note that the doublet and triplet distributions are identical at production, confirming event kinematics are purely a function of chargino mass.  Imposing tracklet selection sculpts this initial $p_T$ spectrum towards the boosted regime, with the lower-lifetime doublets requiring a higher boost in order to survive to the outer pixel layer.  Accounting for the imperfect $p_T$ resolution in reconstructing such short tracklets once again makes these two distinct scenarios indistinguishable.  We will discuss the full impact of this poor $p_T$ resolution when we present our results.

The cut flow for the EW triplet benchmark $(m_{\chi^\pm},\tau_{\chi^\pm}) $ = $(400 \gev, 0.2 \mathrm{\ ns})$ for an integrated luminosity of 36 fb$^{-1}$ can be compared with the corresponding numbers in the ATLAS analysis~\cite{Aaboud:2017mpt} in Table~\ref{tab:sig}.  Requirements below the dashed line are applied to truth charginos with a decay weighting as explained above. Our cut efficiencies are comparable with the ATLAS expected numbers at the event level, but they begin to diverge in tracklet selection due to our implicit assumption that all charginos in selected events are correctly reconstructed as tracklets. Instead, ATLAS find that in the presence of pileup 20\% of charginos are reconstructed including an additional fake hit in the first SCT layer, and hence are vetoed by the disappearance condition (see Figure 3 in ref.~\cite{ATLAS:2019pjd} and accompanying text).  This is reassuringly consistent with the enhancement we observe in the overall number of tracklets passing our selection, and we account for this effect by normalizing our validation results by a factor of 0.8.  The inefficiency mentioned above is mitigated in subsequent versions of the ATLAS tracklet reconstruction algorithm as discussed in Section~\ref{sec:newsearch}.
\begin{table}[th]
  \centering
\begin{tabu}[t]{@{}c c c c@{}}
 \toprule 
Selection requirement                    & This work & ATLAS expected~\cite{Aaboud:2017mpt} &  Ratio \\
\midrule
Lepton veto                              & 1171.8   & 1178  & 1.0 \\
$E_T^\mathrm{miss}$ and jet requirements &  605.9      & 579.1 & 1.0 \\
\tabucline[0.4pt off 2pt]{-}
$p_T,\,\eta$ and quality requirements       & 32.7   & 29.6  &  0.9 \\
Disappearance condition                  & 29.7  & 24.1   & 0.8 \\
\bottomrule
 \end{tabu}
\caption{Cut flow for the ATLAS EW triplet benchmark point, $(m_{\chi^\pm},\tau_{\chi^\pm}) = (400 \gev, 0.2 \mathrm{\ ns})$, for simulated events with an integrated luminosity of $36.1 \mathrm{~fb}^{-1}$, and corresponding ATLAS expectation.  Requirements below the dashed line are applied to truth charginos in our analysis, as explained in the text.}
  \label{tab:sig}
\end{table}

\subsection{Background Modeling}

Modeling the backgrounds to this search is very challenging, not just in principle, as it requires a precise quantitative knowledge of the properties of pile-up, the material budget of the detector, the rates for hard and multiple scatterings inside this material and the ATLAS tracklet reconstruction algorithm, but also in practice because for these processes to combine in a way that yields a fake tracklet is rare.  Indeed ATLAS uses data-driven methods to extract the measured backgrounds from regions of phase space that are signal-poor, such as the low-$\met$ region, and subsequently extrapolates them to the signal region.

With a view to studying the sensitivity of the analysis due to modifications of the selection criteria, we generate at LO the SM processes that contribute to the background, and normalize them to the corresponding ATLAS background distributions in the signal region. These normalizations, or transfer factors, account for the probability that a light SM particle (or particles) in our background simulation is incorrectly reconstructed as a tracklet.  We provide details of the simulation and calculation of the transfer factors for each of the two main sources of background: the hadron/electron and fake combinatoric backgrounds, in Sections~\ref{sec:hadel} and~\ref{sec:fakebg} below, respectively.

\subsubsection{Hadron / electron background}
\label{sec:hadel}

A primary source of background for this search is SM charged particles, such as leptons or hadrons, that undergo a large scattering between the pixel detector and the SCT.  This can be due to multiple interactions with the detector material, or hard bremsstrahlung for leptons. ATLAS combine the hadron and electron components, which are observed to have similar $p_T$ spectra in both the low- and high-$\met$ regions, into a single template~\cite{Aaboud:2017mpt}. We expect this observation to be robust to changes in the tracklet definition, and hence we can model their combined distribution using only the SM single-electron background.   The muon contribution, which has a $p_T$ spectrum that is distinct to that from electron/hadrons, is significantly smaller (see Figs 7 of Ref.~\cite{Aaboud:2017mpt}), and we neglect it from here on.

Electrons are produced with sufficient $\met$ in SM $W+j$, as well as semi-electronic $t\bar{t}$ final states.  We simulate these processes at LO, merged and matched up to 2 additional partons, using the same pipeline as for signal including Delphes detector simulation.  We set the CKKW-L merging scales for $\left(W\rightarrow e\nu_e\right)+j$ and $t\bar t$ to 30 and $45 \gev$ respectively and normalize the resulting distributions to the cross-section at higher order in QCD using a $p_T$-independent K-factor extracted as follows.  For $V+j$ we use Figure 18 of Ref.~\cite{Lindert:2017olm} to extract an approximate ratio of the cross section at NNLO to that at LO, resulting in a K-factor of $1.05/0.7=1.5$.  We normalize the $t\bar{t}$ events to the NNLO+NNLL QCD-corrected total cross section of 819 pb, as calculated in ref.~\cite{Czakon:2011xx} using the Top++2.0 program .  NNLO corrections to the shape of the electron $p_T$ spectrum in both backgrounds was quantified for $V+j$ in~\cite{Lindert:2017olm}, and results in an electron that is up to 15\% harder than the NLO distribution at $p_T=3\tev$. We expect effects of this order to be obscured after $p_T$ smearing and neglect higher-order shape corrections throughout.

We select background events containing exactly one reconstructed electron, but otherwise satisfying the event selection; we treat this electron as our `tracklet', subtracting its reco-level $p_T$ from the reconstructed $\met$ of the event.  We then smear the reco $p_T$ using the smearing function Eq.~\ref{equ:ptsmearing} with parameters $\alpha=1.67,\,\beta=-1.72\tev^{-1},\,\sigma=13.2\tev^{-1}$ (see ref.~\cite{Aaboud:2017mpt}), before applying the tracklet $p_T$, $\eta$ and isolation cuts.  Finally we compute the bin-by-bin transfer factors required to normalize the resultant smeared `tracklet' $p_T$ distribution to ATLAS electron-hadron background data in the high-$\met$ region (Figure 7(c) of ref.~\cite{Aaboud:2017mpt}) using published HEPData~\cite{HEPData}.  The transfer factors, which range from $\mathcal O(10^{-2}) - \mathcal O(10^{-5})$, are found to be consistent with the corresponding factors computed in the low-$\met$ region.  Whilst we would expect the hard scattering probability to scale with the electron $p_T$, note that due to lack of access to the truth momentum spectrum of the measured hadron/electron background, the transfer factors we comput correspond rather to the probability of scattering for an electron given {\it smeared} $p_T$.

\subsubsection{Fake Tracklet Background}
\label{sec:fakebg}
The remaining background is due to mis-reconstructed `fake' tracklets, constituting spatially proximate pixel hits from two or more different particles, likely originating from soft pile-up jets. There are several distinct SM contributions to this background component, which must contain $\met$ and no reconstructed leptons.  The only process with zero leptons and real $\met$ is $\left(Z\rightarrow \nu \nu\right)+j$. We add to this SM processes with unidentified/unreconstructed leptons and $\met$, of which the leading ones are leptonic $W+j$ and semileptonic $t\bar t$ with misidentified or out-of-acceptance leptons of all three flavors. Finally we consider processes with zero leptons, and fake $\met$ arising from jet mismeasurement, the dominant source being SM dijet production.

Since the reconstructed `tracklet' here is not a real object, we cannot simulate the $p_T$ spectrum of the fake background, but instead simply match the summed cross section of the background constituents satisfying event selection to the measured (fitted) fake tracklet cross section in the high-$\met$ region as before.  Later we will use this normalization factor to estimate the scaling of the fake backgrounds as we modify the tracklet definition used in the original analysis.

We simulate $Z+j$ using the signal pipeline, merged and matched up to 3 additional partons with a matching scale of $30 \gev$\footnote{The original rationale behind including a $3^{\rm rd}$ parton in the matching of the $Z+j$ background was to have a more precise modeling of the jet $p_T$ spectrum for the fake tracklet background simulation. However, as described in section~\ref{sec:3layerhadel}, we do not in fact rely on this information in modeling this background.}.  As in the $W+j$ events we apply a flat K-factor of 1.5, read from Ref.~\cite{Lindert:2017olm}, to the matched cross section in order to normalize it to its NNLO QCD value.  To determine the contribution from unreconstructed leptons, we recycle the  $(W\rightarrow e \nu_e)+j$ and semi-electronic $t\bar t$ events generated for the electron/hadron background estimation, including only those events containing an out-of-acceptance or misidentified electron.  We then scale for flavor as follows: events with mis-identified electrons are scaled by 1.2 to account for electronic tau decays; events with out-of-acceptance electrons are scaled by a factor of 3 to include out-of-acceptance muons and taus.

We generate the dijet sample loosely following the Pythia-only method detailed in ref.~\cite{ATLAS:2019mra}, with $p_T$-binned generation for a dijet invariant $\hat{p}_T>45\gev$, with the LO matrix element interfaced to a $p_T$-ordered parton shower.  We use the ATLAS A14 central tune with the `NNPDF2.3LO QCD+QED LO alpha\_s(M\_Z) = 0.119' PDF set.  To maximize high-$p_T$ statistics we oversample events at scale $\hat{p}_T$ by a factor $(\hat{p}_T/10\gev)^4$, and we switch multi-parton interactions off in the interest of speed.  We run the resulting samples through the Delphes ATLAS detector simulation as before.  This gives rise to fake $\met$ in the events at reco-level, due to detector inefficiencies and mismeasurement effects as modeled in Delphes.  To parametrize potential systematic uncertainties in this modeling we introduce an independent normalization factor for dijets in the computation of transfer factors for fake background, Eq.~\ref{eq:eps}.  This factor can also absorb any rescaling of the dijet total cross section due to higher order effects.  Higher order shape effects are irrelevant here since we do not make use of the generated $p_T$ spectra.
The transfer factor, $\epsilon$, can be thought of as the probability to have four pixel hits from two or more charged particles align to fake a tracklet, multiplied by the tracklet reconstruction efficiency.  We assume this is constant across all background components (modulo the factor $f$ parametrizing our ignorance of the fake $\met$ distribution as discussed above) and is approximately independent of kinematics.  The latter assumption was independently verified using the three $\met$ regions of the latest tracklet analysis, ref.~\cite{ATLAS:2022rme}.
\begin{eqnarray}
&\epsilon (\mathcal N^{\mathrm{low} \met}_Z +  \mathcal N^{\mathrm{low} \met}_W +  \mathcal N^{\mathrm{low} \met}_{t \bar t} 
+  f \mathcal N^{\mathrm{low} \met}_{2j}) =  \mathcal N^{\mathrm{low} \met}_\mathrm{ATLAS} \\ 
&\epsilon (\mathcal N^{\mathrm{high} \met}_Z +  \mathcal N^{\mathrm{high} \met}_W +  \mathcal N^{\mathrm{high} \met}_{t \bar t} 
+  f \mathcal N^{\mathrm{high} \met}_{2j}) =  \mathcal N^{\mathrm{high} \met}_\mathrm{ATLAS}.
\label{eq:eps}
\end{eqnarray}
Solving equations Eq.~\ref{eq:eps} yields a fake tracklet reconstruction probability of $\epsilon \approx 10^{-6}$, with a dijet $\met$ rescaling factor $f=0.2$.

\begin{figure}
	\begin{center}
		\includegraphics[width=0.65\textwidth]{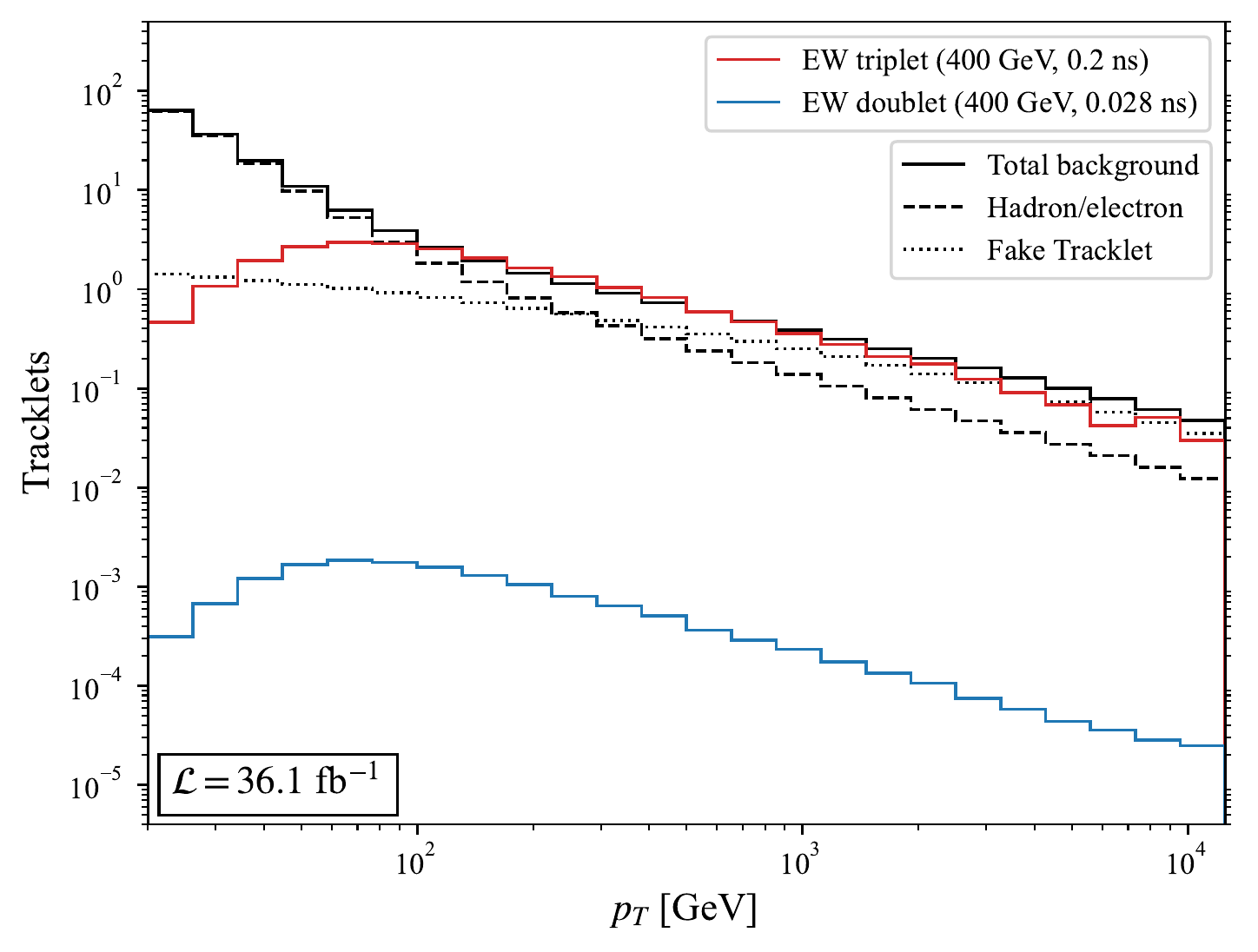}
		\caption{Tracklet $p_T$ distributions normalized to a total luminosity of 36.1 fb$^{-1}$ for EW triplet $(m_{\chi^\pm},\tau_{\chi^\pm}) =(400 \gev, 0.2~\mathrm{ns})$, EW doublet $(m_{\chi^\pm},\tau_{\chi^\pm}) =(400 \gev, 0.028~\mathrm{ns})$ and backgrounds by category, in the high-$\met$ region.}
		\label{fig:sigpT}
	\end{center}
\end{figure}

With all normalizations in place the total number of SM background events, organized by category, process and phase space region, is shown in table~\ref{tab:bkgs}, and the resultant $p_T$ spectrum in the high-$\met$ region can be compared with that of signal events in figure~\ref{fig:sigpT}. 
\begin{table}[htb]
  \centering
\renewcommand{\arraystretch}{1.2}
\begin{tabular}[t]{@{}p{1.8cm}lcc@{}}
 \toprule 
Category &Background process & low  $E^\mathrm{miss}_T$ region  & high  $E^\mathrm{miss}_T$ region    \\
 \midrule
 \multirow{3}{*}{\parbox[r]{1mm}{Electron/ \\hadron}}
        & $W\rightarrow e\nu_e$ & 41.7& 130.2\\
  & semi-electronic $t\bar t$  & 4.6 & 10.2\\
  & {\bf Total} & {\bf 45.2}& {\bf 140.5}\\
 \midrule
\multirow{5}{*}{\parbox[t]{1mm}{ Fake \\tracklets}}
& $Z\rightarrow\nu_\ell\nu_\ell$  & 1.3 & 3.4  \\
& $W\rightarrow l\nu_\ell$ ($l$ not identified) & 2.7 & 5.2  \\
& semi-leptonic $t\bar t$ ($l$ not identified) & 0.2 & 0.4  \\
& Dijets  & 13.1 & 3.5  \\
& {\bf Total} & {\bf 17.3} & {\bf 12.5}\\
\bottomrule
\end{tabular}
\caption{Number of background events normalized to a luminosity of 36.1~fb$^{-1}$, broken down by constituent SM process and phase space region.  The total number of events for each background category and phase space region is identical to that in the ATLAS analysis by construction.
\label{tab:bkgs} }
\end{table}
We see that the background distribution, which is identical to ATLAS' by construction, is hadron/electron-dominated at low tracklet $p_T$ but fake-dominated at high tracklet $p_T$.  Note that although the imperfect tracklet resolution obscures nearly all traces of the large chargino boost, some information remains, resulting in signal distributions that peak around the 100 GeV scale.  Both background distributions, by contrast, are scale-less, decreasing monotonically with tracklet $p_T$.  This allows us to enhance the signal purity using a tracklet $p_T$ cut.  The optimal sensitivity in this analysis is obtained for $p_T>100\gev$, and we display the event distribution in this signal region in table~\ref{tab:SRnumbers}.  We observe agreement between our triplet numbers and ATLAS numbers at $\mathcal{O}(10\%)$, with any differences due to small discrepancies in the reconstructed signal $p_T$ spectrum.  

\begin{table}
\centering
\renewcommand{\arraystretch}{1.2}
\begin{tabular}[t]{@{}r l  l  l@{}}
\cmidrule[0.9pt]{3-4}
&   &\parbox[t]{3cm}{\# events in\\ signal region}  & $p$-value\\
\cmidrule{2-4}
\multirow{3}{*}{Signal $\Bigg{\{}$} &EW triplet $(m_{\chi^\pm},\tau_{\chi^\pm}) =(400 \gev, 0.2~\mathrm{ns})$  & 12.1 & 0.003 \\
&ATLAS expected (triplet)& 13.5                        & 0.001  \\
& \tabdashline
& EW doublet $(m_{\chi^\pm},\tau_{\chi^\pm}) =(400 \gev, 0.028~\mathrm{ns})$ & 0.007 & 0.507  \\
\cmidrule{2-4}
\multirow{3}{*}{Background $\Bigg{\{}$}&Hadron/electron  & 6.1 &   \\
&Fake tracklet & 5.5  &  \\
&{\bf Total} & {\bf 11.6}    &   \\
  \cmidrule[0.9pt]{2-4}
\end{tabular}
\caption{The number of expected events for signal benchmarks and for backgrounds with the tracklet $p_{T} > 100 \gev$ in the signal region. The ATLAS analysis does not list a signal number for the EW doublet. As discussed in sections~\ref{sec:hadel} and \ref{sec:fakebg}, the background normalization and $p_T$ distributions are identical to those in the ATLAS analysis by construction. The $p$-value in the last column is calculated by integrating the probability distribution function for the number of events in the presence of signal past the median value in the absence of signal.}
\label{tab:SRnumbers}
\end{table}
\subsection{Exclusion Sensitivity}
\label{sec:4layercombination}
\begin{figure}
  \centering
  \includegraphics[width=0.85\linewidth]{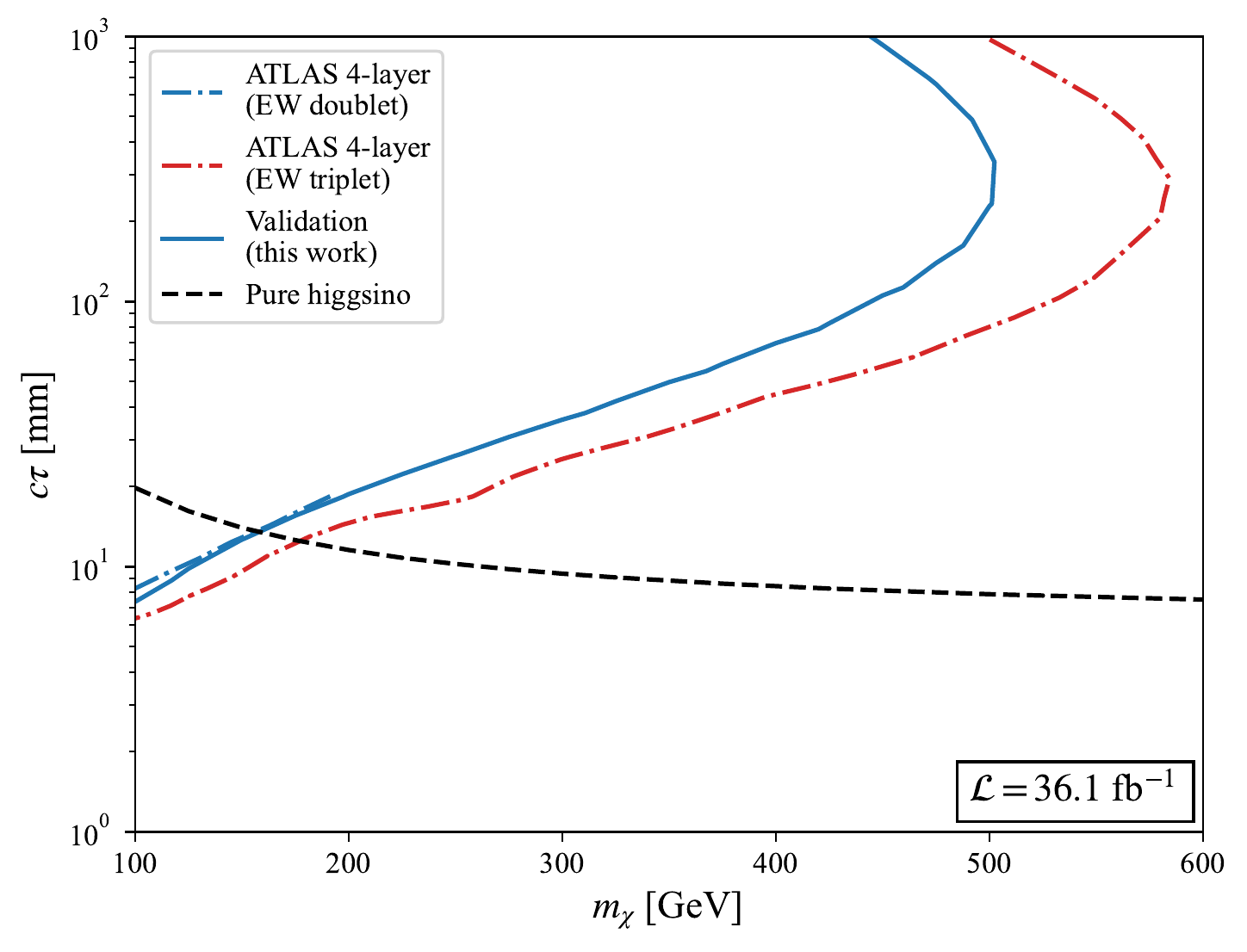}
\caption{The solid blue curve shows our reproduction of the ATLAS exclusion region for the EW doublet (dot-dashed blue curve) using the analysis from ref.~\cite{Aaboud:2017mpt} with an integrated luminosity of 36.1 fb$^{-1}$.  The corresponding ATLAS limit for the EW triplet (dot-dashed red curve) is shown for comparison.
  }
    \label{fig:atlas}
\end{figure}

With our samples and method validated, we proceed to use them to estimate the 95\% exclusion limit in the ATLAS search for EW doublets in the $(m_\chi,c\tau)$ plane, for an integrated luminosity of 36.1 fb$^{-1}$.  Rather than scanning over both mass and lifetime, we simply manually reweight the doublet sample at each mass by the decay probability corresponding to the chosen lifetime, keeping the splitting and decay mode fixed.  We will discuss the validity and ramifications of this assumption in \nameref{sec:conclusions}.
We compare our limit with the corresponding higgsino limit due to ATLAS, as given in figure~\ref{fig:atlas} of ref.~\cite{ATLAS:2022rme} , and see good agreement in the region of the parameter space constrained. The ATLAS limit for the EW triplet is shown for reference.  However it is evident from table \ref{tab:SRnumbers} that applying the triplet analysis to the pure doublet case is ineffectual, with the number of tracklets in the signal region being three orders of magnitude smaller than that for the triplet. While this is partly due to the doublet production cross section, which is roughly a factor of four smaller than the triplet, the root of the problem lies in the shorter lifetime of the doublet at a given mass, which results in an exponential suppression of the number of tracklets satisfying the selection criteria.  We will explore modifications of the analysis that target the shorter doublet lifetime in Section \ref{sec:newsearch}.

\section{Modified search for maximal sensitivity to an EW doublet}
\label{sec:newsearch}

Having validated our simulation of the signal and background by reproducing the ATLAS results, we now move to the main focus of our paper, namely improving the sensitivity of tracklet searches for shorter chargino lifetimes. The main idea of this proposal is to extend the tracklet definition to include also candidates consisting of hits in the inner three pixel layers only, thus including in the search charginos that decay before reaching the fourth pixel layer.  We start by extrapolating the signal and background with the relaxed tracklet selection requirements, without making any changes to the event selection. Not surprisingly, we find that this results in a significant increase in the background, and we then consider how each of the two main backgrounds can be further reduced.

\subsection{Signal yield for a 3-layer search}

\begin{figure}
  \centering
  \includegraphics[width=0.6\linewidth]{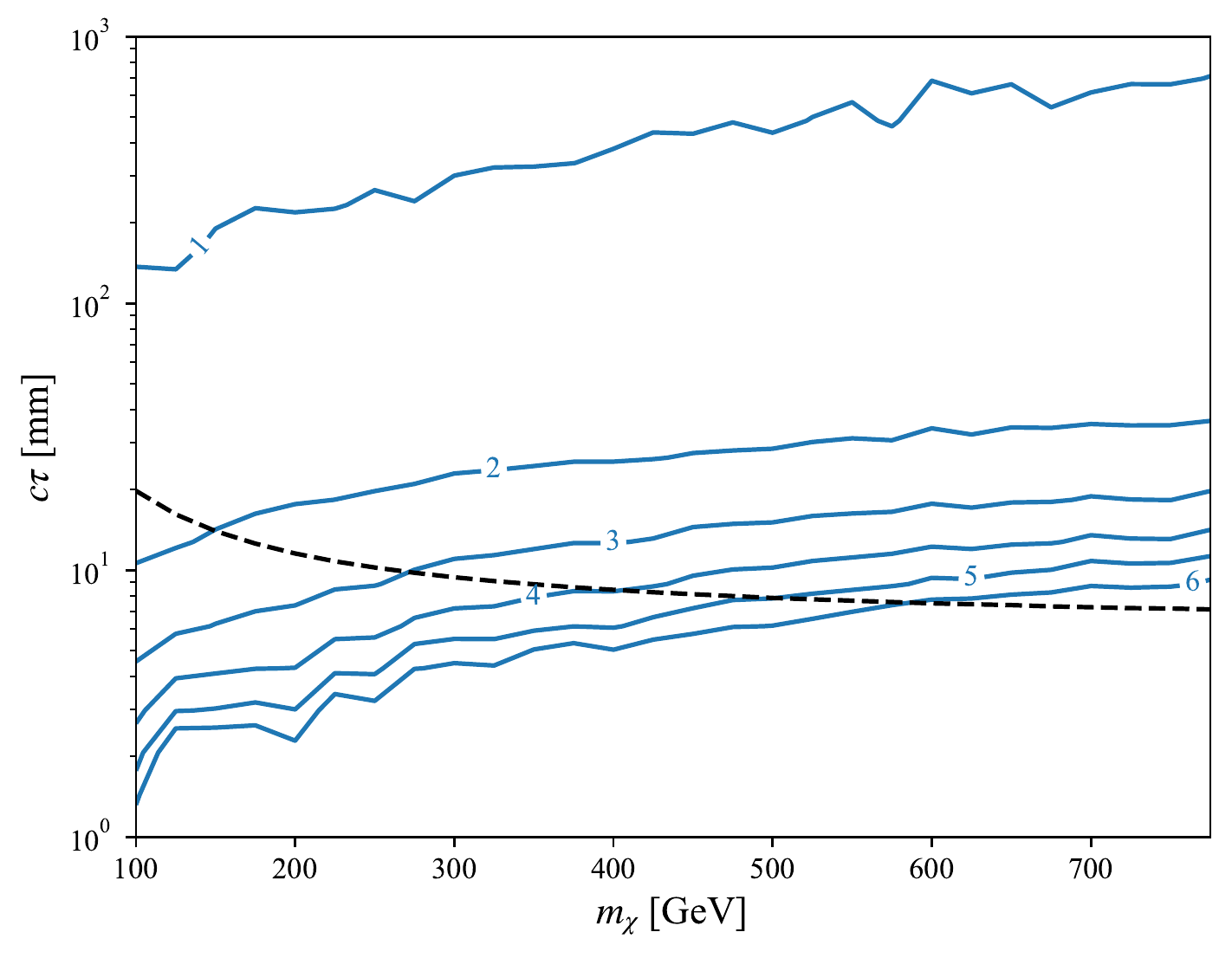}
  \caption{Contours of signal enhancement for a 3-layer tracklet selection as compared with a conventional 4-layer selection, showing a two- to six-fold increase in the number of pure higgsinos passing the event and tracklet selection of the modified analysis in the parameter space of interest at the LHC.  Contours above 6 are omitted due to inefficiencies in generation. The dashed line represents the pure higgsino decay lengths w.r.t its mass.}
\label{fig:signalimprovement}
\end{figure}

Relaxing the tracklet requirement to include those consisting of hits on only the inner three pixel layers gives rise to a significant enhancement of the signal yield.  This enhancement can be seen in figure~\ref{fig:signalimprovement}, where the contours denote the ratio of the tracklet selection efficiency for 3-layer tracklets to that for 4-layer tracklets, and would result in a two- to six-fold increase in the number of pure higgsinos passing event and tracklet selection (table \ref{tab:EventSelection}), in the parameter space of interest at the LHC.  However as noted above this signal enhancement is accompanied by an enhancement in the background, and a detailed analysis of the signal spectrum and corresponding background growth and spectrum is necessary to assess the potential of this search strategy. 

En route to determining the signal spectrum there are two outstanding issues: first, how the reconstruction efficiency for the 3-layer tracklet differs from the 4-layer one, and second, how the shorter length of the tracklet affects its $p_T$ resolution.  Fortunately these questions have been addressed by ATLAS in their 2019 tracking and vertexing study, ref.~\cite{ATLAS:2019pjd}.  We see in figure 4a that the 3-layer reconstruction efficiency is roughly independent of chargino decay radius prior to reaching the SCT, and we apply a flat factor of 0.92 to our gen-level charginos to account for this.  Note that this increase in reconstruction efficiency as compared with the original analysis is due to an updated reconstruction procedure which filters out fake hits on the SCT that were previously being associated with a chargino tracklet, causing the tracklet to be vetoed by the disappearance condition.  As regards the degradation of the $p_T$ resolution with decreasing tracklet length, ATLAS note that the resolution scales with $1/L^2$, where $L$ is the transverse length of the tracklet.  We would thus expect the resolution of the 3-layer tracklet to be a factor $(122.5 \,\textrm{mm}/88.5 \,\textrm{mm})^2\sim 2$ worse than a 4-layer one, corresponding to the squared ratio of pixel layer positions, and we scale the $\sigma$ parameter in eq.~\ref{equ:ptsmearing} accordingly. We do this uniformly for all signal events, even though some signal events may contain a fourth pixel hit.

\subsection{Extrapolating the hadron/electron background}
\label{sec:3layerhadel}

Recall that tracklets can result from SM hadron/electron background due to hard or multiple scatterings of charged particles off of the material in the inner detector.  Whilst the scattering must occur between the outermost pixel layer and the SCT in order to be mistaken for a 4-layer tracklet, for the 3-layer search any scattering between the third pixel and first SCT layers could result in an imitation tracklet. Note that we are using an inclusive definition of 3-layer tracklets for the background just as with signal, namely a fourth pixel hit is allowed in the tracklet definition. We can estimate the relative probability of a background event satisfying the 3-layer and 4-layer tracklet selection criteria by taking the ratio of the material budget, as measured using hadronic conversion candidates in Figure 11 of ref.~\cite{ATLAS:2017oro}, integrated over the corresponding intervals. Explicitly, we rescale the 4-layer hadron/electron background by an overall normalization factor
\[
 \frac{\int_{88.5\,\textrm{mm}}^{299.0\,\textrm{mm}}\textrm{(conversion candidates)}\,d\kern 0.03em r}{\int_{122.5\,\textrm{mm}}^{299.0\,\textrm{mm}}\textrm{(conversion candidates)}\,d\kern 0.03em r}=2.14\,
 \]
where $r$ is the transverse distance from the beamline.
In their updated analysis, ref.~\cite{ATLAS:2022rme}, ATLAS use a `calorimeter veto' to suppress the hadron/electron background, filtering out tracklet candidates with more than 5 GeV of energy in calorimeter deposits within $\Delta R < 0.2$ of the tracklet direction. This results in a reduction of the background by a factor of $8.5\times 10^{-2}$, the ratio of hadron/electron tracklets with associated calorimeter energy less than 5 GeV to those with calorimeter energy greater than 5 GeV, as read from Figure 2 of the article.  We assume that the distribution shown in the figure, corresponding to 4-hit tracklets in the control region $\met > 100\gev$, tracklet $p_T < 60\gev$, can be extrapolated unchanged to 3-hit tracklets in the signal region of our analysis, and impose this suppression factor on our rescaled 3-layer hadron/electron background.  The effect of the calorimeter veto on the fake background and signal is minimal, resulting in a reduction in both by under 5\%, and we neglect this small effect from here on.

 We also need to account for the degradation of the reconstructed $p_T$ for the shorter hadron/electron tracklet.  Our procedure is similar to that implemented for signal tracklets, where we smear the tracklet $p_T$ using Eq.~\ref{equ:ptsmearing} with resolution $\sigma$ scaled by the squared inverse length ratio $(122.5/88.5)^2$. We apply this smearing to the reconstructed electrons in our SM samples, and then rescale them by the transfer factors computed in section~\ref{sec:hadel}, in order to account for the probability of faking a tracklet. As with the signal, we do this for all background events, even though a fraction of background events will leave a fourth pixel hit.

\subsection{Extrapolating the fake tracklet background}
\label{sec:3layerfake}

Since the fake tracklet background is due to unrelated hits in the pixel detector being mis-reconstructed as belonging to a single tracklet, we would expect it to be significantly enhanced if we include 3-layer tracklets in the tracklet definition.  Below, we describe how we estimate the size of this background in our proposed search, and in the following subsection, how it may be reduced.

We use a simple geometric simulation of fake tracklet generation and reconstruction in order to estimate the scaling of the fake background with tracklet length.  We consider the pixel layers in an $r$-$z$ slice of the pixel detector (corresponding to a $\Delta \phi$ section) and generate a fixed number of randomly distributed `hits' in the layers, representing `unassociated' hits left over after the standard tracking algorithm has been run. We then run a simplified tracklet reconstruction algorithm (described below) on this pseudo-data and identify tracklet candidates. Whilst we do not expect this simplified procedure to reproduce all the subtleties of a realistic detector and reconstruction procedure, we intend it to capture most of the relevant physics of fake tracklets, and use the ratio of 3- and 4-layer tracklet candidates in our simulation to estimate the scaling of the fake background in going from the 4-layer analysis to the 3-layer one. It may be possible to use geometric arguments and the conjecture that the fake tracklets arise from unassociated pixel hits to derive a scaling behavior for the $p_T$ spectrum, but we will not attempt this here. We take the $p_T$ spectrum for fake tracklets in a 3-layer search to be the same as in a 4-layer search.

\begin{figure}
  \centering
  \includegraphics[width=0.5\linewidth]{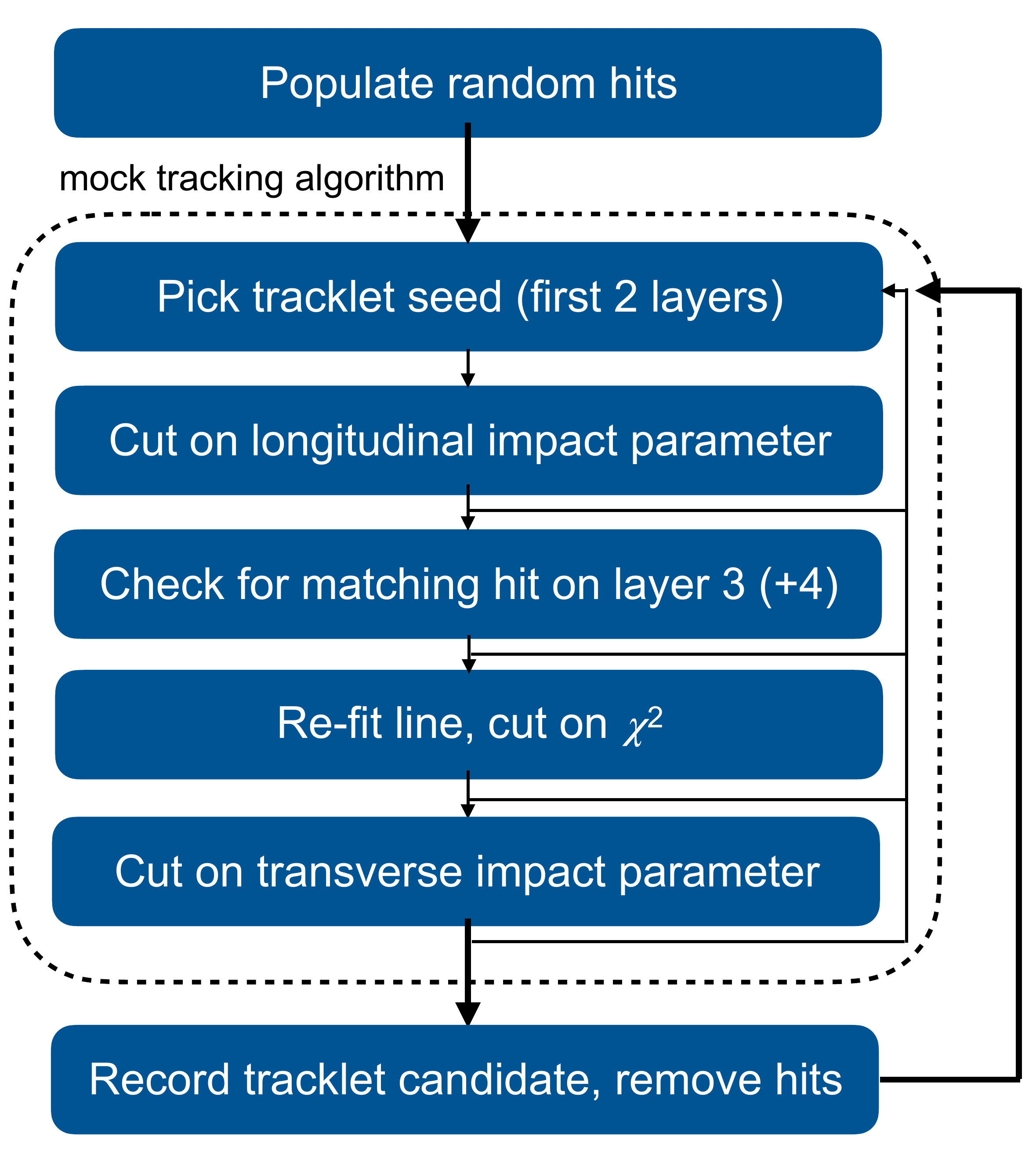}
\caption{A visual representation of our simplified tracklet reconstruction algorithm.}
\label{fig:trackingsim}
\end{figure}

The details of our simplified detector modeling and reconstruction algorithm, chosen to closely mimic the ATLAS procedure~\cite{ATLAS:2019pjd}
, are as follows (see figure~\ref{fig:trackingsim} for a visual representation):
First we randomly generate a fixed number of `hits' to represent the unassociated hits on the first pixel layer.  The number of hits is scaled for subsequent layers, with two possible scalings: a constant scaling, and a $1/r$ scaling. The latter is a strawman for modeling the loss of lower-$p_T$ particles having circular trajectories with radii smaller than the radius of outer layers.
We use pairs of hits, one on the innermost layer, one on the next layer, as tracklet seeds and we construct lines passing through any pair of such points. We demand that the line have a longitudinal impact parameter $|z_0|< 10~\mathrm{mm}$ along the beam axis from the hard vertex, which we take to be fixed at $r=z=0$. For each such tracklet candidate, we look for additional hits (on layer 3 for a 3-layer tracklet candidate and one each on layers 3 and 4 for a 4-layer tracklet candidate) within $\Delta z = 0.060~\mathrm{mm}$ (corresponding to the physical pixel size) of the line going through the seed hits. Once a 3- or 4-layer tracklet candidate is identified that satisfies this criterion, we re-fit the line, using all hits of the tracklet candidate (imposing $\chi^2<2.0$, normalized for $\chi^2$ with 1 or 2 degrees of freedom, respectively), and we further demand that the re-fitted line have a transverse impact parameter ($d_0 <0.5~\mathrm{mm}$). The cutoff for $\chi^2$ is chosen to stabilize the result, so that increasing the cutoff further does not result in a significant increase in the number of tracklet candidates. Once a tracklet candidate satisfying all criteria is found, the corresponding hits are removed, and the procedure is repeated until no more tracklet candidates remain.

\begin{figure}
  \centering
  \includegraphics[width=0.45\linewidth]{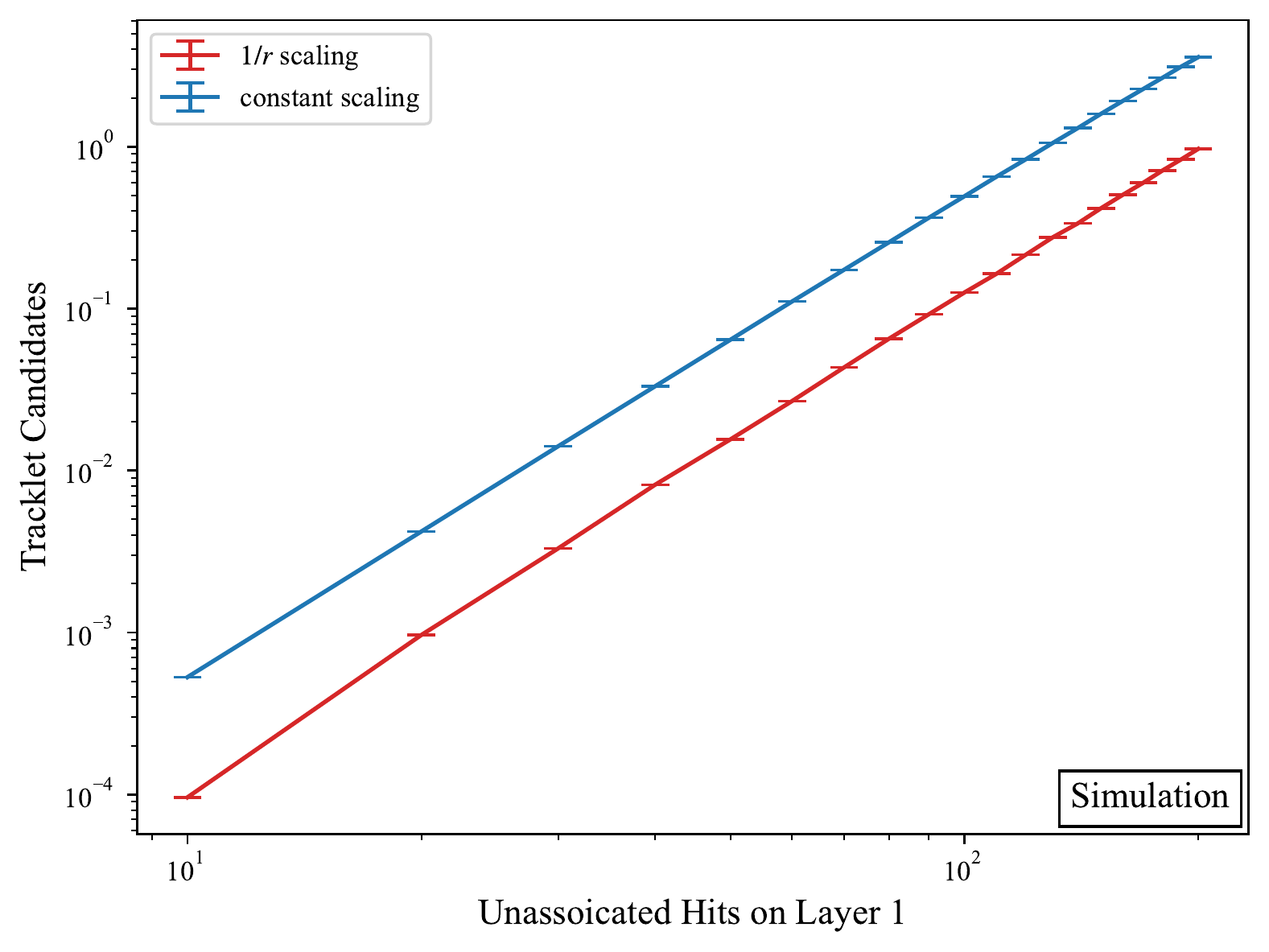}
  \includegraphics[width=0.45\linewidth]{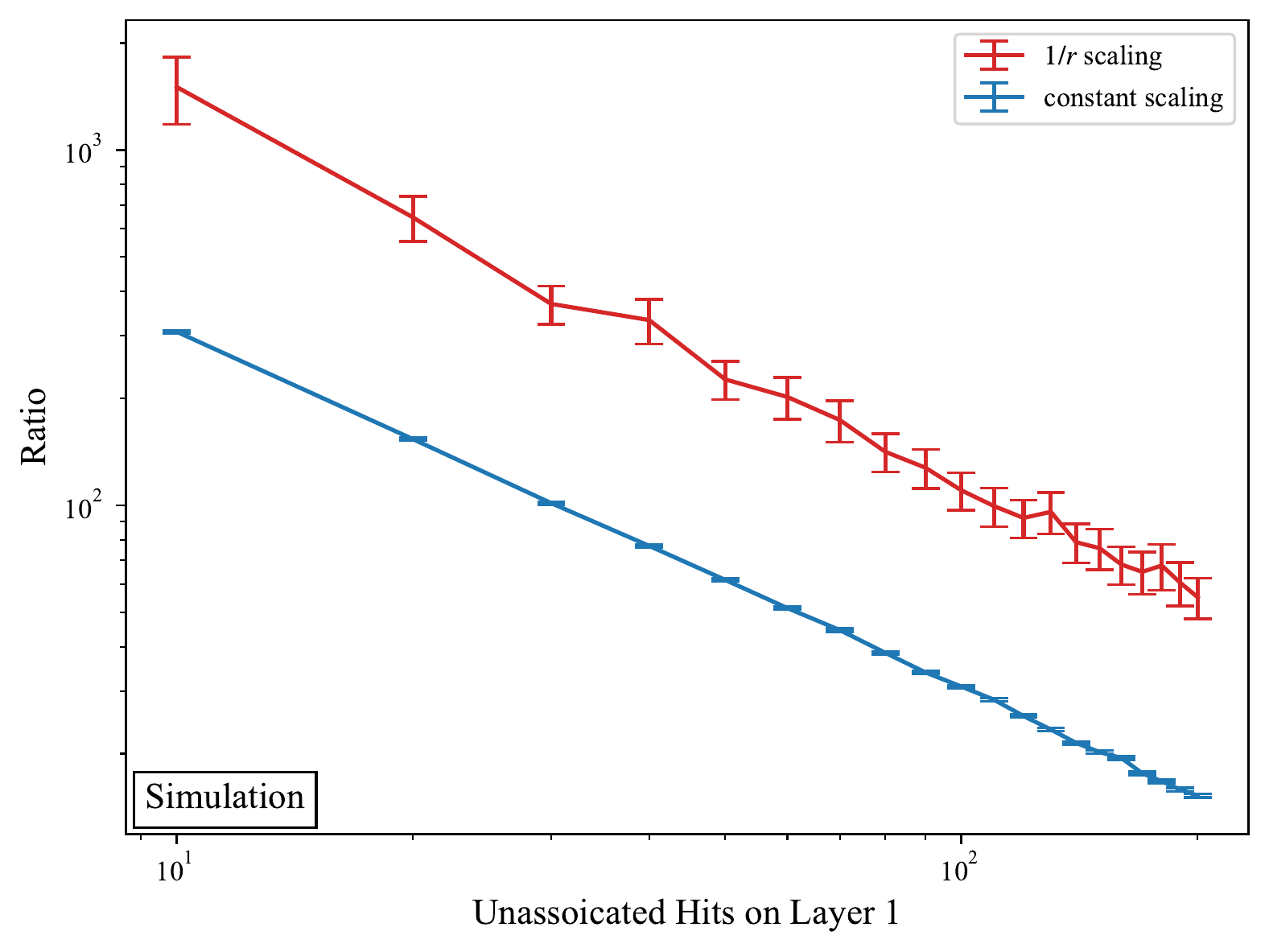}
\caption{{\it Left:} The number of 3-layer tracklet candidates, up to an unknown normalization factor, as a function of the number of unassociated hits on the innermost pixel layer.  {\it Right:} The enhancement in the number of tracklet candidates in going from a 4-layer search to a 3-layer search.  Results were obtained from the simulation described in the main text, for a constant and $1/r$ scaling of the number of hits on the first pixel layer, in going to outer detector layers.  The  arbitrary normalization in the left plot encodes fine details of pile-up distribution, detector layout and tracklet reconstruction that are not included in our simulation.  We expect these unknowns to fall out in the ratio.}
\label{fig:recoresults}
\end{figure}

In figure~\ref{fig:recoresults}, we plot the results of our simulation, which demonstrate a robust power-law behaviour with varying numbers of unassociated hits on the innermost pixel layer.
In the left panel, we show the number of 3-layer tracklet candidates, up to a scaling factor for physics not captured in our simulation, for two scalings of the number of hits on the outer pixel layers, as described above. In the right panel, we plot the ratio of 3-layer tracklet candidates to the number of 4-layer tracklet candidates, which we expect is independent of the unknown normalization factor.  Note that this ratio grows rapidly with decreasing number of hits as expected, and also with $1/r$ scaling, since longer tracklets are harder to reconstruct with fewer hits.  We adopt a benchmark value of 50 unassociated hits on the innermost layer, and we use the corresponding ratios to rescale the 4-layer tracklet background measured by ATLAS to obtain a 3-layer fake tracklet background estimate.

\begin{figure}[htb]
  \centering
  \includegraphics[width=0.83\linewidth]{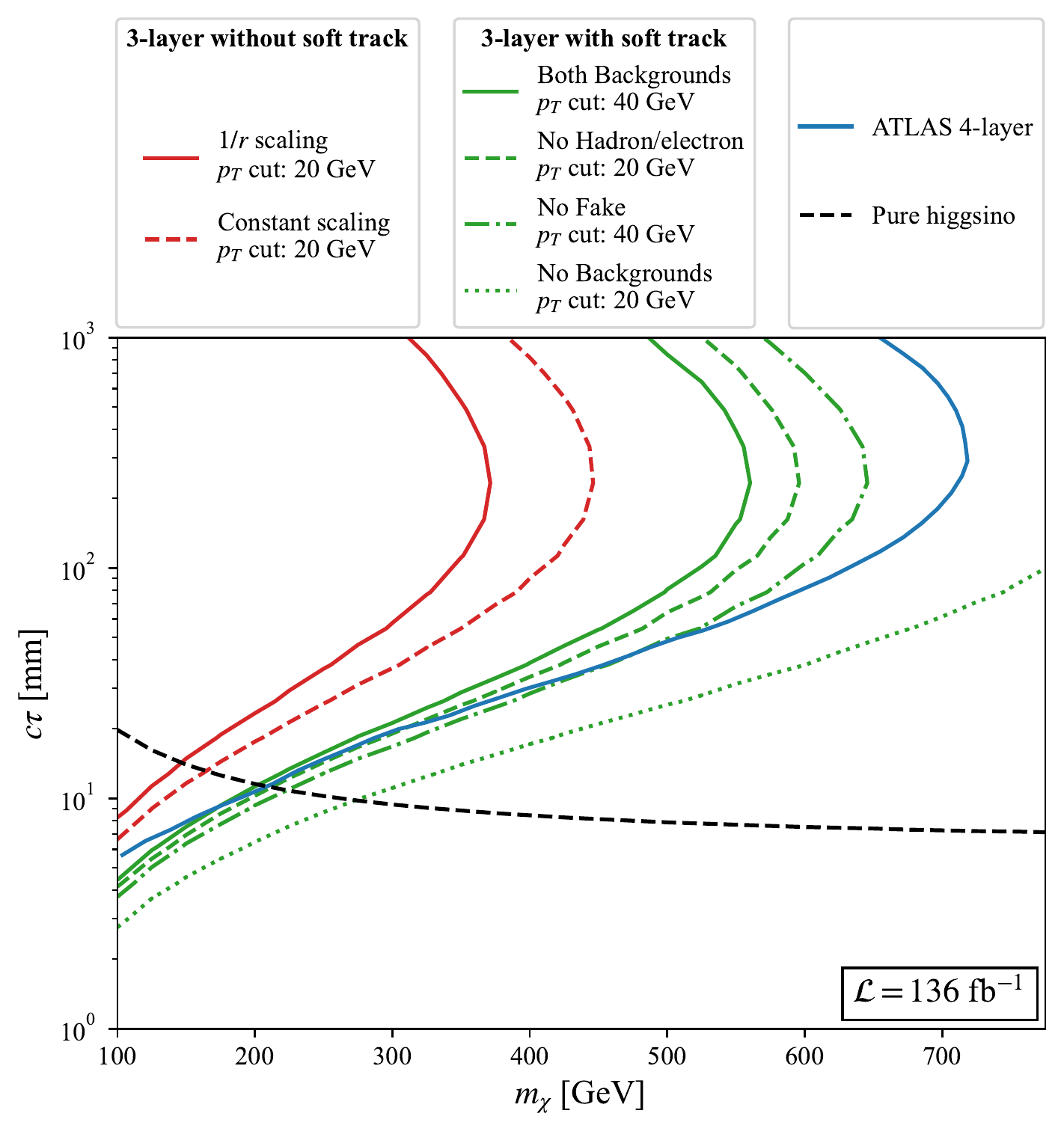}
  \caption{The sensitivity in the $(m_{\chi},c\mkern 1mu\tau)$ plane of our proposed tracklet searches for an EW doublet fermion, for an integrated luminosity of 136~fb$^{-1}$. The blue curve corresponds to the doublet reach of the 136 fb$^{-1}$ ATLAS 4-layer search, ref.~\cite{ATLAS:2022rme} . The two red curves correspond to the pure 3-layer search with conservative and agggressive scalings for the fake tracklet background as described in section~\ref{sec:3layerfake}. The green curves correspond to a 3-layer search with soft track requirement as described in section~\ref{sec:3layerfakeplussoft}, with the different line styles corresponding to different background benchmarks.  For the dotted green curve we assume all backgrounds are eliminated by the soft track requirement; for the dot-dashed green curve the hadron/electron background is left unchanged and the fake tracklet background is eliminated;  for the dashed green curve, the hadron-electron background is eliminated and the fake tracklet background is reduced to its level in the 4-layer search; for the solid green curve the hadron/electron background and fake backgrounds are retained at their levels in the 3-layer and 4-layer searches, respectively.  The values of the tracklet $p_T$ cut, chosen in each case for maximal sensitivity along the (black dashed) pure EW doublet curve, are listed in the legend.}
\label{fig:bgvariation_reach}
\end{figure}

Clearly, this sharp increase in the fake tracklet background has a negative impact on the sensitivity of the search. This can be seen in figure~\ref{fig:bgvariation_reach}, where the blue curve represents the reach of the ATLAS 4-layer analysis, with tracklet $p_T$ cut optimized for maximal sensitivity along the (black dashed) EW doublet curve, as $p_T>40\gev$. The red curves represent the reach 95\% exclusion limit of the 3-layer analysis with the two benchmarks for the fake tracklet background, and with the tracklet $p_T$ cut optimized at 20~GeV. We see that the orders-of-magnitude increase in the fake background overwhelms the signal enhancement from going to shorter tracklets, resulting in a degradation of the search sensitivity as compared with the 4-layer analysis.  We now turn our attention to how the 3-layer search can be modified to reduce the fake tracklet background.

\subsection{Adding a soft track requirement}
\label{sec:3layerfakeplussoft}
We will exploit the presence of the soft pion from the chargino decay in selecting tracklet candidates in order to reduce the fake tracklet background. Tracking and vertexing efficiencies for soft track reconstruction were studied by ATLAS in ref.~\cite{ATLAS:2019pjd}; we use their work to guide us in estimating the efficiency changes due to this selection criterion.

\begin{figure}[htb]
  \centering
  \includegraphics[width=0.6\linewidth]{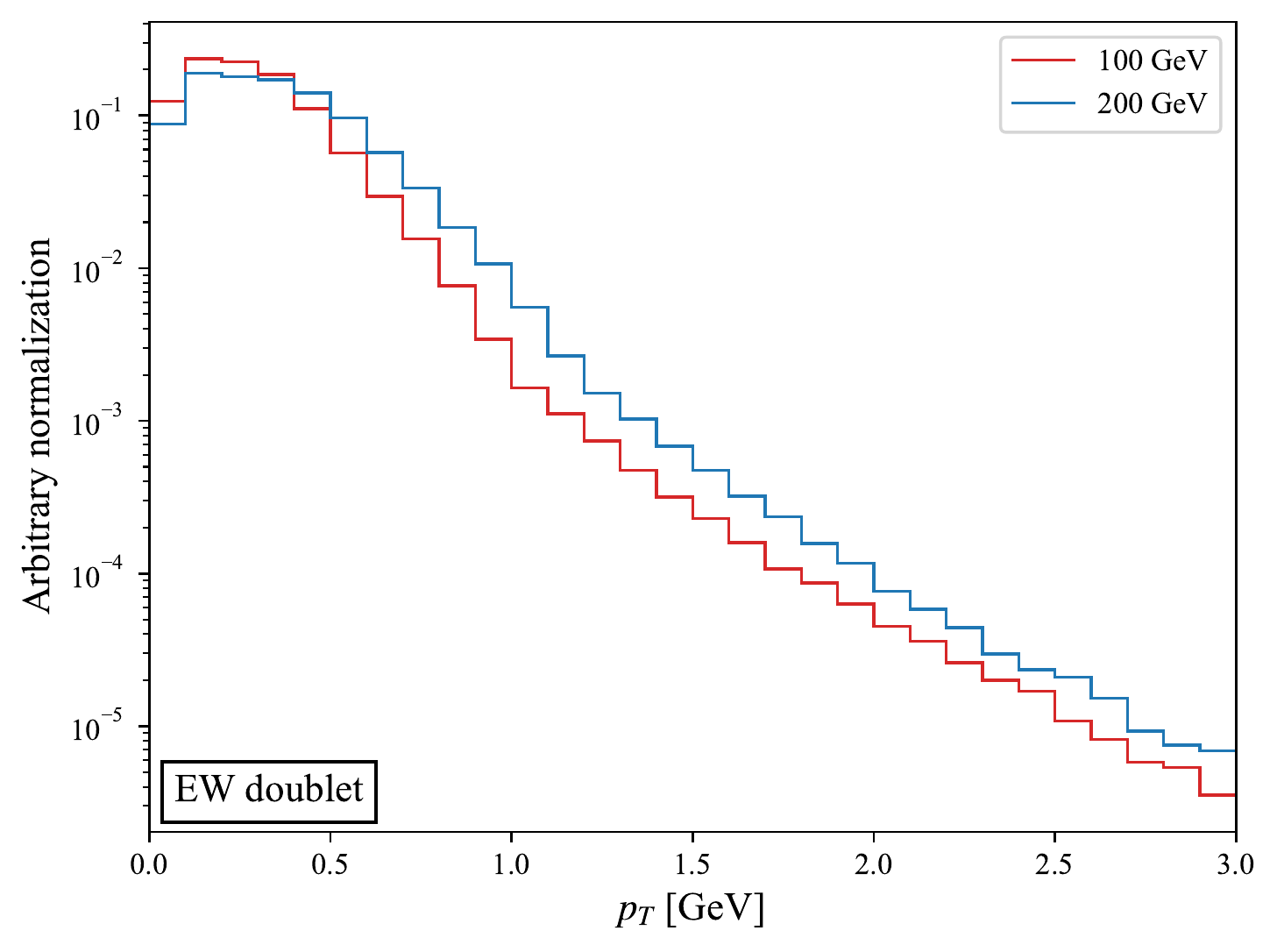}
\caption{The truth $p_T$ distribution of soft pions produced in the chargino decay, for two pure higgsino signal benchmarks.}
\label{fig:softpionpt}
\end{figure}

In figure~\ref{fig:softpionpt} we show the truth $p_T$ distribution of the soft pion for two pure higgsino benchmark points after tracklet selection.
The small difference in shape is attributable to the larger splitting, and hence shorter lifetime, of the 200-GeV chargino, which requires it to be more boosted in order to satisfy tracklet selection.

In their higgsino soft track study, ref.~\cite{ATLAS:2019pjd}, ATLAS reconstruct soft tracks from the chargino decay using SCT hits lying within a small cone centered on the chargino tracklet direction, with the apex set by the last chargino pixel hit.  Because we lack the tools to do a full detector simulation and tracklet reconstruction, we simply apply to our events the soft track selection efficiency reported by ATLAS, read from figure 7a of ref.~\cite{ATLAS:2019pjd}, which ranges from 0.23 at $p_T = 250~$MeV to 0.75 at $p_T > 1.0$~GeV.  We remind the reader that we are setting the chargino branching ratio to a charged pion, the dominant decay channel for a pure higgsino, equal to 1 for the purposes of this analysis.  In principle the arguments above would also apply to events with subdominant decays, which include decays to electrons and muons, and these would pass soft track selection, but with higher efficiency due to the smaller mass of the charged daughter.

  \begin{table}[htb]
\centering
\renewcommand{\arraystretch}{1.3}
\begin{tabular}[t]{@{}l  c  c@{}}
\cmidrule[0.9pt]{2-3}
 & \multicolumn{2}{c}{Events in signal region}\\
  &\parbox{4cm}{\centering 3 layer\\(tracklet $p_T> 20\gev$) } & \parbox{4cm}{\centering 3 layer plus soft\\ (tracklet $p_T > 40\gev$)}\\
\cmidrule{1-3}
\parbox{5.6cm}{EW doublet \\$(m_{\chi^\pm},\tau_{\chi^\pm}) =(400 \gev, 0.028~\mathrm{ns})$} & 0.094 & 0.044  \\
\tabucline[0.4pt off 2pt]{-}
Hadron/electron  & 115.1 & 10.8\\
Fake tracklet (constant scaling) & 2942.6   &  33.0\\
{\bf Total background} & {\bf 3057.8} & {\bf 43.8}  \\
\tabucline[0.4pt off 2pt]{-}
$S/\sqrt{B}$ & $1.70 \times 10^{-3}$ & $6.48 \times 10^{-3}$\\
  \cmidrule[0.9pt]{1-3}
\end{tabular}
\caption{Number of signal and background events for our benchmark signal point, normalized to a luminosity of 36.1~fb$^{-1}$, for searches proposed in this section.}
\label{tab:3hitnumbers}
\end{table}

While we would expect the soft track requirement to significantly reduce both components of the background to this search, there are no existing studies that allow us to estimate the background efficiency due to the soft track requirement.  We instead bracket each background using benchmarks that range from the conservative to the aggressive, and estimate the reach for each of these cases, with the true exclusion boundary likely lying between the two extremes.
For both backgrounds we use as an aggressive estimate the complete elimination of the background.  On the conservative end we assume no reduction to the hadron/electron background due to the soft track requirement (i.e. it is the same as in the 3-layer analysis), and for the fake tracklet background, that it is reduced to its level in the 4-layer analysis.  Part of the motivation for this latter choice is that it is quantified by ATLAS and requires no additional scaling assumptions.  We record the number of signal and background events populating the signal region for our pure higgsino validation benchmark point, with and without the soft track requirement, in table~\ref{tab:3hitnumbers}.

We display our results as green curves in the $(m_\chi, c\tau)$ plane in figure~\ref{fig:bgvariation_reach}, for an integrated luminosity of 136~fb$^{-1}$. As for our other exclusions, the tracklet $p_T$ cut is optimized along each curve for maximum sensitivity along the (black dashed) pure higgsino line, with the cut values listed in the legend. The solid, dashed, dot-dashed and dotted green curves correspond to different benchmarks for the hadron/electron and fake tracklet backgrounds, as described in the previous paragraph. In order for this result to gain sensitivity with respect to the ATLAS analysis, we see that we must assume some amount of background reduction due to the soft tracklet requirement.  Depending on the level of background suppression achieved, the pure higgsino exclusion for Run 2 can be pushed to somewhere in the mass range $210-270\gev$.

Let us also consider the effect of the tracklet $p_T$ resolution on the sensitivity of tracklet searches. As we have seen, the $p_T$ resolution for 4-layer tracklets is already relatively poor, to the extent that the reconstructed signal $p_T$ spectrum is shaped by the smearing function, rather than the true chargino $p_T$ (see figure~\ref{fig:pT4}).  This results in the distinctive chargino $p_T$ spectrum, sculpted towards high $p_T$ due to the tracklet selection for short lifetime, being obscured and unexploitable. This resolution degrades further for 3-layer tracklets due to their smaller lever arm.  Improving the $p_T$ resolution would significantly increase the sensitivity of tracklet searches, particularly for lower lifetimes.

We illustrate this point in figure~\ref{fig:resvariation_reach}. As in figure~\ref{fig:bgvariation_reach}, the blue curve represents the 95\% exclusion for the ATLAS 4-layer analysis, and the dot-dashed green curve corresponds to the reach of the 3-layer analysis with fake tracklet background assumed set to zero by the soft track requirement, but hadron/electron background unchanged from the 3-layer case. Using the latter curve as a reference, we add two more green curves, a dashed curve where we return to the original 4-layer $p_T$ resolution parameters, and a dotted curve where we assume perfect resolution and use truth $p_T$ spectra, both for the signal and the electron background.  We normalize the integral of the truth electron spectrum to the total number of hadron/electron background events, computed by rescaling the observed ATLAS background by the material budget factor from section~\ref{sec:3layerhadel}.  Reverting to the 4-layer resolution may be achieved in practice by using the reconstructed position of the chargino decay vertex as a fourth hit,  in order to more precisely measure the tracklet curvature, as proposed in ref.~\cite{Fukuda:2017jmk}.  Unsurprisingly, with perfect tracklet $p_T$ resolution, a much higher tracklet $p_T$ cut can be employed without any impact to signal events.  The improvement in the reach of the search is quite dramatic, with a pure higgsino exclusion limit of over 300 GeV.  This surpasses even the limit from our zero background estimate due to its 100\% signal efficiency.  While a perfect tracklet resolution is of course quite unrealistic, the lesson learned here is that any technique for sidestepping the resolution issue and selecting true high-$p_T$ tracklets will have a huge impact on the reach of the search. We will propose a way to achieve this outcome in an upcoming paper.

\subsection{Extrapolating to Run 3}
We project our estimates of the 3-layer reach to the Run 3 luminosity of 300~fb$^{-1}$ by scaling up our signal and background simulations by the ratios of their LO cross sections at 13 and 13.6 TeV center-of-mass (CM) energies, assuming the same transfer factors as computed in sections~\ref{sec:hadel} and \ref{sec:fakebg}.  We further scale up the fake tracklet background due to the increased pile-up.  We assume that the number of unassociated hits scales like the number of charged particles in each bunch crossing, which is in turn proportional to $\langle\mu\rangle$, the average number of collisions per bunch crossing. With this choice, our benchmark number of unassociated hits on layer 1 goes up from 50 in Run 2 to 188 for Run 3, with the number of 4-layer tracklets increasing by a factor of 185 (for either the constant or the $1/r$ scaling), and the number of 3-layer tracklets increasing by an additional factor of 16 for the constant scaling and 62 for the $1/r$ scaling, as can be read off from the right panel of figure~\ref{fig:recoresults}.

\begin{figure}[htb]
  \centering
  \includegraphics[width=0.85\linewidth]{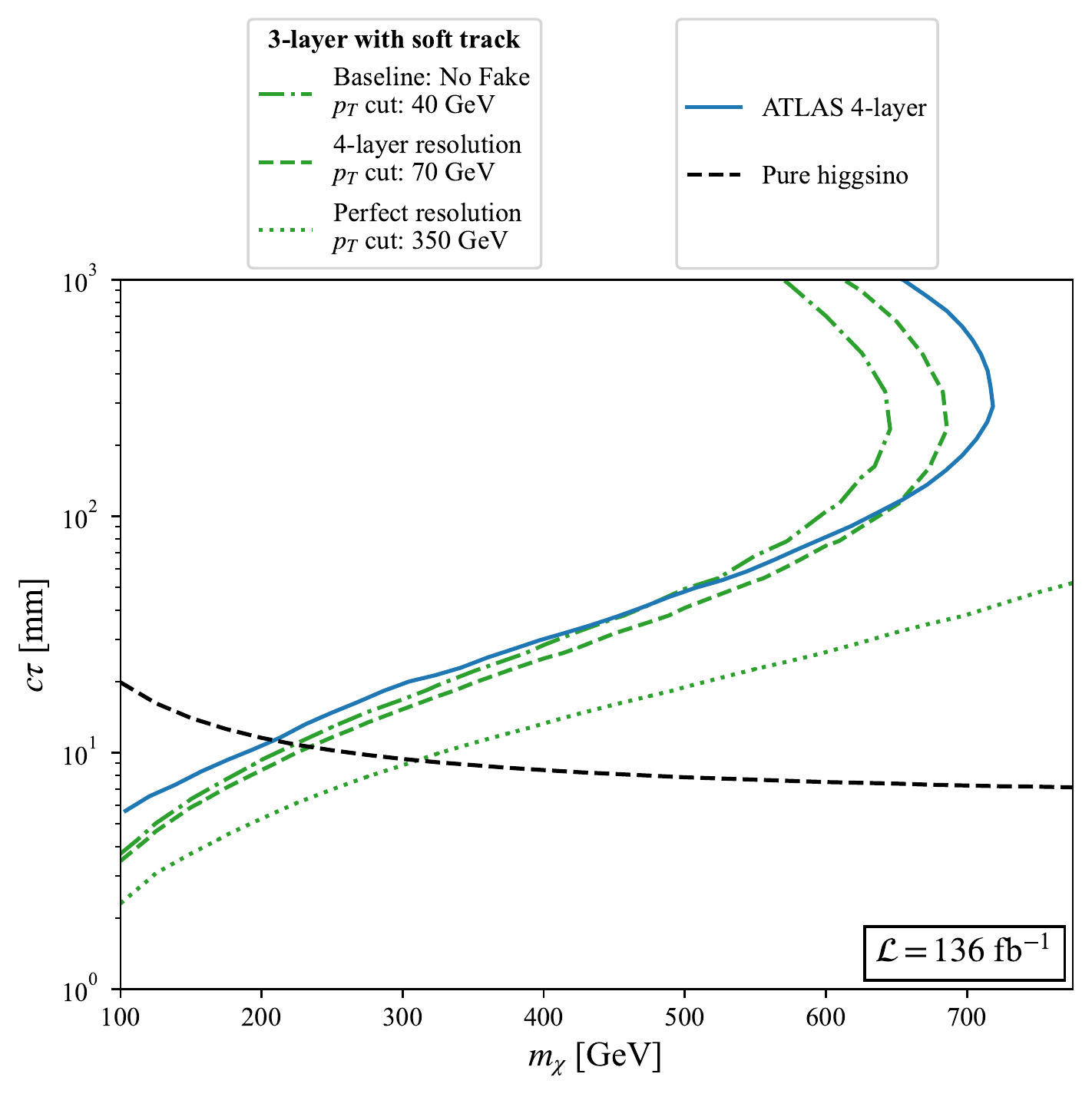}
\caption{We illustrate the effect of the chargino $p_T$ resolution on the reach of the 3-layer tracklet search with a soft track requirement, for an integrated luminosity of 136~fb$^{-1}$. The solid blue and dot-dashed green curves are the same as in figure~\ref{fig:bgvariation_reach}. The dashed green curve is obtained from the dot-dashed green one by reverting to the $p_T$ resolution of 4-layer tracklets, and the dotted green curve is obtained by assuming perfect $p_T$ resolution. The optimal value of the chargino $p_T$ cut used to obtain each curve is listed in the legend.}
\label{fig:resvariation_reach}
\end{figure}

\begin{figure}[htb]
  \centering
  \includegraphics[width=0.48\linewidth]{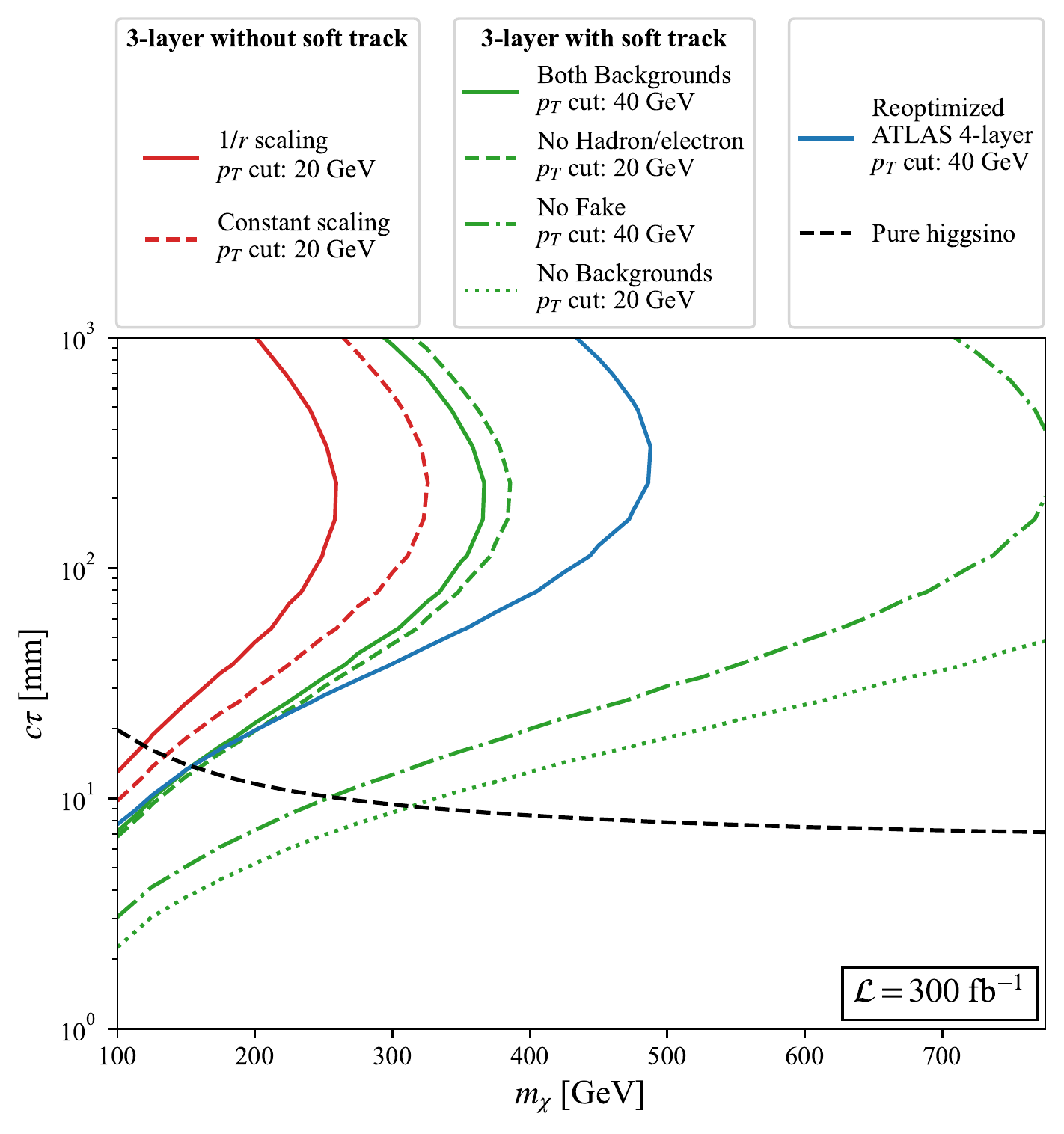}~\includegraphics[width=0.48\linewidth]{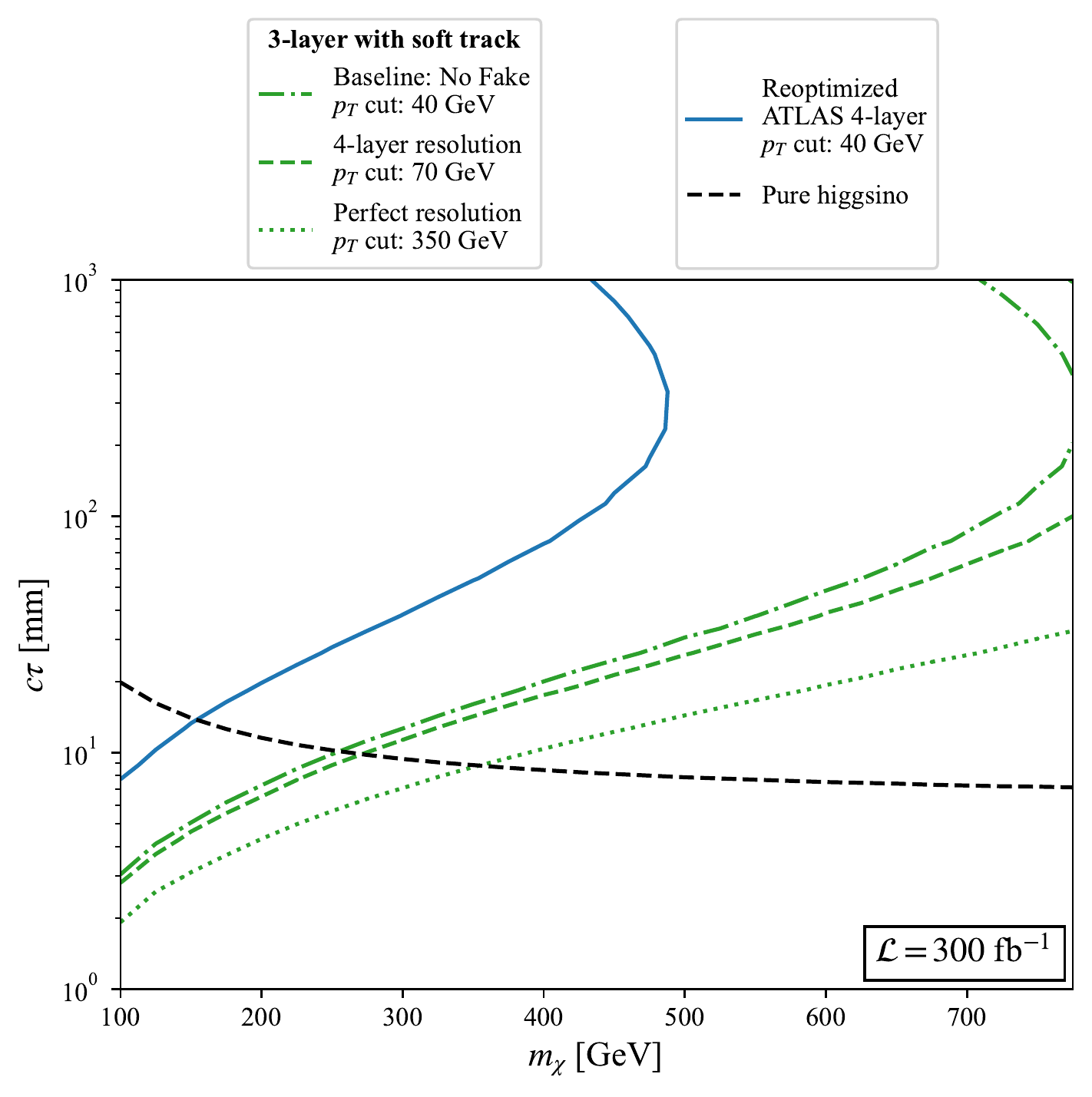}
  \caption{{\it Left:} Projection of figure~\ref{fig:bgvariation_reach} to the LHC Run 3 CM energy and luminosity. {\it Right:} Projection of figure~\ref{fig:resvariation_reach} to the LHC Run 3 CM energy and luminosity. The background benchmarks are chosen as described in the captions of the Run 2 figures.  The increase of the fake tracklet background due to the higher pile-up numbers in Run 3 is described in the main text.}
\label{fig:bgresvariation_reach_run3}
\end{figure}

Although we would also expect the kinematics of signal and background events to change with CM energy, we expect this effect to have small impact on the search reach, for the $0.6$-TeV increase in CM energy.  At any rate, we expect any enhancement to be more pronounced for signal compared to background due to the exponential effect of the chargino boost on the signal yield.  Therefore our projections can be taken to be conservative. We show our projections in the two panels of figure~\ref{fig:bgresvariation_reach_run3}, which are the counterparts of figures~\ref{fig:bgvariation_reach} and \ref{fig:resvariation_reach} for Run 3.  We see that the power of the Run 3 search is driven by the fast growth of the fake tracklet background with the CM energy and luminosity, and any search with the nominal fake background component {\it decreases} in sensitivity as we move to higher luminosity.  By contrast, if the fake background component can be reduced, by imposing a soft track requirement, for instance, or by improving the chargino $p_T$ resolution, the 95\% exclusion limit for the pure higgsino increases to 310 GeV for zero backgrounds, or up to 360 GeV for perfect $p_T$ resolution.

\section{Discussion}
\label{sec:conclusions}

In this work, we have outlined how a tracklet-based search requiring three or more pixel layer hits, combined with selection cuts to reduce backgrounds, most importantly a soft track requirement, can help improve the reach for arguably the most challenging WIMP candidate from a collider physics viewpoint: a Pseudo-Dirac electroweak doublet fermion, which in a supersymmetric setting would correspond to a higgsino-like LSP. We have described how we model the two dominant backgrounds, namely the hadron/electron and fake tracklet backgrounds, and we have reproduced the ATLAS sensitivity in a published tracklet search for validation. We have then extrapolated our signal and backgrounds to include 3-layer tracklets in our selection, including the effect of a soft track requirement to reduce backgrounds. We have studied the 95\% exclusion acheivable with Run 2 data and we have also extrapolated our results to the Run 3 energy and luminosity. Even with conservative assumptions for the background we estimate that the Run 2 higgsino exclusion can be pushed to 210~GeV.  With more aggressive background estimates this number may go up to 270~GeV.

This is a good point to revisit various simplifying assumptions that were made in our analysis and consider their impact on the results.  Within the simplified model defined in Eq.~\ref{eq:simplifiedmodel}, with strictly renormalizable couplings between the EW multiplet fermion and SM states, the relationship between the mass of the neutralino and its mass difference to the chargino, or equivalently the chargino lifetime, is fixed.
In scanning over lifetime in our plots, we are implicitly allowing for the presence of higher-dimensional operators that modify the chargino-neutralino mass splitting, and hence the chargino lifetime, without appreciably changing the normalization or dynamics of the chargino production process.  While this is clearly the case for small deviations in lifetime away from the pure doublet limit, it would be expedient to verify how far one can push in lifetime for a given mass, and remain within a pure doublet simplified model for the purposes of collider phenomenology.  In other words, how large do the UV-suppressed contributions to the splittings need to be in order to appreciably change the doublet lifetime?\footnote{Note that the chargino width is strongly sensitive to the mass difference, scaling like $(\delta m)^3$ for the doublet, so the entire $c\tau$ range shown in our reach plots corresponds to splittings in the interval $0.15\gev \le \delta m \le 0.6\gev$.} This question has been studied in the context of a particular UV completion, the MSSM and its variants.  Within vanilla SUSY a different higgsino lifetime can be achieved by varying the mass of the wino and bino states, allowing them to mix with the higgsino. A potentially important caveat in our analysis is that we keep the chargino-neutralino mass splitting fixed as we scan over the chargino lifetime. The only relevant effect of the mass splitting is that it determines the $p_T$ of the soft decay pion, and hence the signal selection efficiency of the soft track requirement.  Note however that the kinematics of the decay pion is also affected by the parent chargino mass and boost.  We saw that the latter is necessarily large for any charginos passing tracklet selection, which results in a similarly large boost for the daughter.  We expect this to be the dominant effect in setting the kinematics of the decay pion in any parameter space that enjoys increased sensitivity due to the soft track requirement, an expectation that is borne out by the small differences in reconstruction efficiency for the soft track seen in the ATLAS study~\cite{ATLAS:2019pjd} for their two benchmark parameter points.

Finally, we have emphasized that the tracklet $p_T$ resolution is currently a significant limiting factor in the sensitivity of the search. In future work, we will propose a method to improve upon this limitation by using the $dE/dx$ of the tracklet as a more accurate way to determine its $p_T$. Measuring the tracklet momentum by using $dE/dx$ has another important advantage. In Run 3, ATLAS will reduce the $\met$ trigger threshold, which would make it possible to reduce the $\met$ requirement among the selection cuts for a tracklet analysis and increase the signal efficiency. This may appear somewhat counterintuitive, since naively one may expect that the $\met$ would be correlated with the chargino $p_T$, and therefore that the charginos which live long enough to satisfy tracklet selection criteria would occur in events with large $\met$. This is however not the case, since the $\met$, which as defined, correlates with the $p_T$ of the chargino-pair system (or chargino-neutralino system, depending on the production channel), but not with individual chargino momenta. Our signal modeling confirms that a reduced $\met$ cut can result in a significant improvement in the signal efficiency. The price to pay for reducing the $\met$ cut is of course a large increase in the backgrounds. Once again however, a hard cut on the tracklet $p_T$, measured much more accurately by using $dE/dx$ can then be combined with the lower $\met$ requirement to increase signal while keeping the backgrounds under control. We will quantify the improvement in the sensitivity of the search by following this strategy in future work.

\acknowledgments

The work of MG is partially supported by the U.S. Department of Energy, Office of Science, High-Energy Physics, under award number DE-SC-0010107. The research of CK and TY is supported by the National Science Foundation Grant Numbers PHY-1914679 and PHY-2210562. The authors acknowledge the \href{http://www.tacc.utexas.edu}{Texas Advanced Computing Center} (TACC) at The University of Texas at Austin for providing HPC resources that have contributed to the research results reported within this paper.

\section*{Code availability}
All codes written for this analysis will be shared on reasonable request to the corresponding author.

\bibliographystyle{JHEP}
\bibliography{bibtex}{}

\end{document}